\documentclass[12pt,a4paper,final]{iopart}
\usepackage{iopams}
\usepackage{graphicx}
\usepackage[usenames,dvipsnames]{xcolor}
\usepackage{url,cite}
\usepackage{longtable}
\usepackage{tabularx,booktabs}

\usepackage[font=footnotesize,labelfont=bf]{caption}
\graphicspath{ {./} }

\def\lsim{\mathop{\hbox{${\lower3.8pt\hbox{$<$}}\atop{\raise0.2pt\hbox{$\sim$}}
$}}} \def\gsim{\mathop{\hbox{${\lower3.8pt\hbox{$>$}}\atop{\raise0.2pt\hbox{$
\sim$}}$}}} 
 3
 
\newcommand\eprint[1]{\href{http://arXiv.org/abs/#1}{#1}}
\definecolor{ballblue}{rgb}{0.0,0.3,.79}
\definecolor{dgreen}{rgb}{.1,.6,.1}
\definecolor{MyDarkRed}{rgb}{0.71,0.14,0.07}
\definecolor{MyDarkBlue}{rgb}{0.1,0,0.7}
\definecolor{charcoal}{rgb}{0.21, 0.27, 0.31}
\definecolor{goldenbrown}{rgb}{0.6, 0.4, 0.08}
\definecolor{lincolngreen}{rgb}{0.11, 0.35, 0.02}
\usepackage{hyperref}
\usepackage{array}
\hypersetup{
    colorlinks=true,
    citecolor=blue,
    linkcolor=magenta,
    filecolor=cyan,      
    urlcolor=lincolngreen,
}

\begin{document}
\setcounter{footnote}{0}
\title[Temperature gradient as signature of a multiply connected Universe]{The variance of the CMB temperature gradient: \\a new signature of a multiply connected Universe}
\author{Ralf Aurich$^{\,1}$, Thomas Buchert$^{\,2}$, Martin J. France$^{\,2}$\\ and Frank Steiner$^{\,1,2}$}
\smallskip
\address{$^1$Ulm University, Institute of Theoretical Physics, \\$\ \, $Albert-Einstein-Allee 11, D--89069 Ulm, Germany

$^2$Univ Lyon, Ens de Lyon, Univ Lyon1, CNRS, \\$\ \, $Centre de Recherche Astrophysique de Lyon UMR5574, F--69007, Lyon, France
\\
Emails: ralf.aurich@uni--ulm.de, buchert@ens--lyon.fr, martin.france@ens--lyon.fr, frank.steiner@uni-ulm.de}
\begin{abstract}
In this work we investigate the standard deviation of the Cosmic Microwave Background (CMB) temperature gradient field as a signature for a multiply connected nature of the Universe. CMB simulations of a spatially  infinite universe model within the paradigm of the standard cosmological model present non-zero two--point correlations at any angular scale. This is in contradiction with the extreme suppression of correlations at scales above $60^{\circ}$ in the observed CMB maps. Universe models with spatially multiply connected topology contain typically a discrete spectrum of the Laplacian with a specific wave-length cut-off and thus lead to a suppression of the correlations at large angular scales, as observed in the CMB (in general there can be also an additional continuous spectrum). Among the simplest examples are $3-$dimensional tori which possess only a discrete spectrum. To date, the universe models with non-trivial topology such as the toroidal space are the only models that possess a two--point correlation function showing a similar behaviour as the one derived from the observed Planck CMB maps. In this work it is shown that the normalized standard deviation of the CMB temperature gradient field does hierarchically detect the change in size of the cubic $3-$torus, if the volume of the Universe is smaller than $\simeq 2.5 \cdot 10^3$ Gpc$^3$. It is also shown that the variance of the temperature gradient of the Planck maps is consistent with the median value of simulations within the standard cosmological model.
All flat tori are globally homogeneous, but are globally anisotropic. However, this study also presents a test showing a level of homogeneity and isotropy of all the CMB map ensembles for the different torus sizes considered that are nearly at the same weak level of anisotropy revealed by the CMB in the standard cosmological model.
\end{abstract}

\clearpage
\section{Introduction}
\label{introduction}

An important open problem in cosmology is the fundamental question whether our Universe is spatially infinite or finite.
This question about the Universe at large is concerned with the global geometry and topology of the Universe.
Modern physical cosmology is based on the theory of General Relativity.
The Einstein field equations are, however, differential equations and thus determine only the local physics but not the global geometry and topology.
Therefore, at present, the only possibility to decide about the global and large-scale properties of the Universe consists in comparing predictions of different models (in the framework of General Relativity) with observational data.

Important clues about the early Universe, its large-scale structure and time evolution are provided by the temperature fluctuations (anisotropies) $\delta T$ of the Cosmic Microwave Background (CMB). 

The CMB was discovered in 1965 by Penzias and Wilson \cite{penzias1} 
in a study of noise backgrounds in a radio telescope as a nearly
isotropic radiation with antenna temperature (3.5$\pm1.0$)K at a wave-length of $7.5\,$cm.\footnote{
``In between the star $\zeta$ Oph (Ophiuchi) and the earth there is a cloud of cold molecular gas, whose absorption of light produces dark lines in the spectrum of the star. In 1941, Adams \cite{adams},
following a suggestion of McKellar, found two dark lines in the spectrum of $\zeta$ Oph that could be identified as due to absorption of light by cyanogen (CN) in the molecular cloud ... From this, McKellar concluded \cite{mckellar1} that a fraction of the CN molecules in the cloud were in the first excited rotational
component of the vibrational ground state, ... and from this fraction he estimated an equivalent molecular temperature of $2.3\,$K. Of course, he did not know that the CN molecules were
being excited by radiation, much less by black-body radiation.
After the discovery by Penzias and Wilson, several astrophysicists
independently noted that the old Adams-McKellar result could be
explained by radiation with a black-body temperature at wavelength
$0.264\,$cm in the neighborhood of $3\,$K.'' (Citation from \cite{weinberg}.)} 
But from this they could not conclude that they were observing a black-body spectrum as predicted by the Big-Bang paradigm. It was $25$ years later that the satellite COBE 
(Cosmic Microwave Background Explorer, active life-time 1989-1993) \cite{mather1} could show with the FIRAS instrument that the CMB data follow an almost perfect Planck spectrum with 
the present mean value of the CMB temperature $T_{\hbox{\scriptsize CMB}}$=(2.735$\pm$0.060)K. Nine years later, COBE finally obtained $T_{\hbox{\scriptsize CMB}}$=(2.725$\pm$0.002)K \cite{mather2} (see section \ref{sec:definition} for the present best value).
In 1992, COBE discovered with the DMR instrument the CMB temperature anisotropies\cite{smoot1, wright1, hinshaw1}, subsequently measured with more and more precision by the
space probe missions WMAP (Wilkinson Microwave Anisotropy Probe, active life-time 2001-2010) \cite{bennett1, hinshaw2, spergel1, bennett2} and Planck (Planck probe active life-time 2009-2013)
\cite{planck1, planck2, planck3, planck4, planck5, planck6}.

A basic quantity characterizing the anisotropies of the CMB and probing the primordial seeds for structure formation is the full-sky two-point correlation function (hereafter 2--pcf) of the temperature 
fluctuation $\delta T({\boldsymbol{\hat n}})$, observed for our actual sky in a direction given by the unit vector ${\boldsymbol{\hat n}}$, defined by  
\begin{equation}
\label{equ:2--pcf}
C^{\rm obs}(\vartheta):=\left<\delta T({\boldsymbol{\hat n}})\delta T({\boldsymbol{\hat n'}}) \right>\, \quad {\rm with} \quad {\boldsymbol{\hat n}}\cdot{\boldsymbol{\hat n'}} = \cos \vartheta \ ,
\end{equation}
where the brackets denote averaging over all directions $\boldsymbol{\hat n}$ and $\boldsymbol{\hat n'}$ (or pixel pairs) on the full sky that are separated by an angle $\vartheta$. 
Since $C^{\rm obs}(\vartheta)$ corresponds to \textit{one} observation of the actual CMB sky from our particular position in the Universe, the average in equation~\eref{equ:2--pcf} should not be confused with an ensemble average. The ensemble average could be either an average of the observations from every vantage point throughout the Universe, or the average of an ensemble of realizations of the CMB sky in a given cosmological model (see section~\ref{sec:definition}).

$C^{\rm obs} (\vartheta)$ has been measured for the first time in 1992 by COBE \cite{smoot1, wright1} from the 1--year maps, and in 1996 from the 4--year maps \cite{hinshaw1}.
The COBE data revealed small correlations in the large angular range $\Upsilon$ delimited by $70^{\circ} \le \vartheta \le$ $150^{\circ}$ which later has been confirmed with high precision by 
WMAP \cite{bennett1, hinshaw2, spergel1, bennett2} and Planck \cite{planck1, planck2, planck3, planck4, planck5, planck6}.  
COBE compared the observed correlation functions with a large variety of theoretical predictions within the class of FLRW (Friedmann-Lema\^\i tre-Robertson-Walker) cosmologies, 
including flat and non-zero constant-curvature models with radiation, massive and massless neutrinos, baryonic matter, cold dark matter (CDM), 
and a cosmological constant $\Lambda$, using both adiabatic and isocurvature initial conditions, see e.g. \cite{bond1, holtzman1}. 
From COBE observations it was concluded \cite{smoot1} that the two-point correlations, including the observed small values of $C^{\rm obs}(\vartheta)$ in the range $\Upsilon$, 
are in accord with scale-invariant primordial fluctuations (Harrison-Zel'dovich spectrum with spectral index $n=1$) and a Gaussian distribution as predicted by models of inflationary cosmology.
Thus, there was no indication that the small correlations measured in the angular range $\Upsilon$ could hint to a serious problem, or even to new physics. 
The situation changed drastically with the release of the first-year WMAP observations that will be discussed below. 

At this point it is worth to mention that at the time of COBE, i.e. before 1998, the Hubble constant was not well-determined (the uncertainty amounting to a factor of $2$ or more);
the acceleration of the time-evolution of the scale-factor of a FLRW cosmology \cite{riess1, perlmutter1} was not yet discovered and thus the value of the cosmological constant was not known.
Also the low quadrupole was already clearly seen by COBE, but was usually dismissed due to cosmic variance or foreground contamination. 

COBE observations were used in 1993 in an attempt \cite{stevens1, stabo} to detect CMB temperature fluctuations specific of the discrete spectrum of metric perturbations of a Universe with $3-$torus topology.
And several other authors emphasized that the COBE observations might hint to a non-trivial topology of our Universe and called this field of research Cosmic Topology \cite{lachiezerey1}.
Another signature of multiply connected universes on the CMB, based on the identified circles principle was proposed since 1996 \cite{cornish1, cornish2} and observational analyses of
the COBE data were made using this principle \cite{roukema1, roukema2}.   

The first-year data by WMAP led to today's standard model of cosmology
\cite{bennett1, hinshaw2, spergel1}, a spatially flat $\Lambda-$dominated universe model seeded by nearly scale-invariant adiabatic Gaussian fluctuations, the $\Lambda$CDM model with cold dark matter and a 
positive cosmological constant $\Lambda$. The fact that the non-Gaussianities of the primordial gravitational fluctuations are very small is nicely confirmed by the recent Planck data \cite{buchert1}.
\begin{figure}[!htb]
\includegraphics[height=0.6\textheight,width=0.6\textwidth, angle=270]{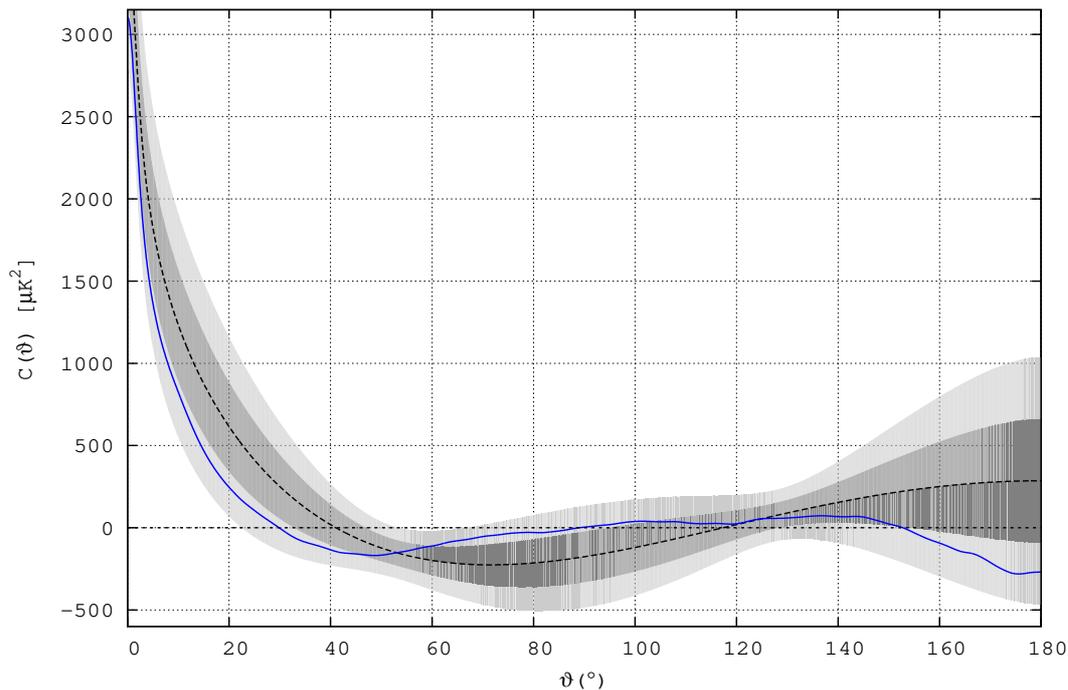}
\caption{The average two-point correlation function of $100\,000$ CMB simulation maps without mask in the infinite $\Lambda$CDM model according to Planck 2015 \cite{planck3} cosmological 
parameters (in black dash line), $\pm 1\sigma$ in the dark shaded area, and $\pm 2\sigma$ in the light shaded area ($68$ and $95$ percent confidence levels, respectively).
This is compared to the average two-point correlation function of the four foreground corrected Planck maps, NILC, SEVEM, SMICA and Commander-Ruler (NSSC) in solid line (blue for the online version).
}
\label{fig1}
\end{figure}

There remain, however, intriguing discrepancies between predictions of the $\Lambda$CDM model and CMB observations: one of them is the lack of any correlated signal on angular scales greater than
60$^{\circ}$ \cite{spergel1}, \cite{aurich7}, \cite{aurichsteinerthen}, \cite{aurich1}, \cite{aurich8}, \cite{aurich2}, \cite{spergel2}, \cite{aurich10}, \cite{roukema3}, \cite{copi1}, \cite{aurich11}, \cite{aurich13}, \cite{aurich12}, \cite{bernui1}. Further anomalies are e.g. the low quadrupole and a strange alignment of the quadrupole with the octopole \cite{tegmark1, deoliveira1, schwarz1}.
These anomalies were still questioned on the basis of the seven-year WMAP data \cite{bennett2}, and it is only with the 
sharper spatial and thermal resolution of Planck that their existence in the CMB data have a robust statistical standing \cite{planck1}. 
The observed severe suppression of correlations at large scales does not appear in the simulated sky map examples of the CMB in a $\Lambda$CDM model.
Figure \ref{fig1} shows the average 2--pcf of the four Planck foreground corrected CMB observation maps without mask, NILC, SEVEM, SMICA and Commander-Ruler 
(their ensemble hereafter named NSSC) compared to the average 2--pcf of one hundred thousand $\Lambda$CDM CMB maps at a resolution $N_{\hbox{\scriptsize side}} =128$, with $l_{\hbox{\scriptsize max}}=256$ 
and a Gaussian smearing (defined in equation \eref{equ:gausskernel}) of $2^{\circ}$ (full width at half maximum). The calculation of the 2--pcfs is made in the spherical harmonic space imposing 
isotropy and homogeneity for the Planck NSSC 2--pcf which shows no correlation between $80^{\circ}$ and $150^{\circ}$. Also the 2--pcf average behaviour of the $\Lambda$CDM ensemble differs strongly from the one of the 
CMB observation maps by WMAP and Planck. 
Approximately $0.025 \%$ of the $\Lambda$CDM realizations have a 2--pcf displaying the same large-angles suppression as the $5-$year WMAP map \cite{copi1}.\\

A further discrepancy occurs on scales below $\vartheta\approx 50^{\circ}$, where the $\Lambda$CDM simulations also reveal, on average, larger correlations than those observed by WMAP and Planck 
(see figure \ref{fig1} and, e.g., figure 3 in \cite{schwarz1}).
The angular range $\vartheta \le 50^{\circ}$ of the 2--pcf depends on all multipoles ($l \ge 2$) of the observed power spectrum (see e.g. the Planck spectrum, figure 57) \cite{planck3})
in the case of no or very small smoothing (see equation \eref{equ:gausskernel}). There is a large contribution from the first acoustic peak and also from the higher peak structure which appears
up to the large $l'$s (i.e. the smallest angles limited by the instrument resolution). Note, however, that the very large multipoles ($l \ge 900$) are strongly suppressed by Silk damping.  
The `high' multipole moments ($l\ge30$) do not differ very much for $\Lambda$CDM and the `topological' models, the crucial contribution to $C(\vartheta)$, which leads to the discrepancy for 
$\vartheta \le 50^{\circ}$, comes from the `low' multipoles (mainly for $l \le 29$) where the power spectrum shows a lack of power for the quadrupole and a characteristic `zig-zag structure'
(see, e.g., \cite{spergel1}, \cite{aurich7}, \cite{aurichsteinerthen}, \cite{aurich1}, \cite{aurich8}, \cite{aurich2}, \cite{spergel2}, \cite{aurich10}, \cite{copi1}, \cite{aurich11}, \cite{aurich13}, \cite{aurich12}, \cite{bernui1}).

In addition, the 2--pcf $C(\vartheta)$ of the $\Lambda$CDM model reveals a negative dip between $50^{\circ}$ and $100^{\circ}$ and a positive slope beyond and up to $180^{\circ}$.
Thus, on average, these CMB sample maps for an isotropic and homogeneous infinite $\Lambda$CDM model display a non zero 2--pcf for any separation angle $\vartheta$ except those in 
the two narrow regions of cancellation around $40^{\circ}$ and $120^{\circ}$. For the observed CMB by Planck, WMAP or COBE the lack of correlations at large angular scales finds a natural explanation 
in cosmic topology: compared to the CMB simulation maps of the $\Lambda$CDM model in an infinite Universe, the suppression of the 2--pcf at large angular scales of the Planck CMB maps is consistent with finite spatial 
sections of the Universe.

Before 1998 there was the theoretical prejudice that the Universe is flat (total density parameter $\Omega_{\textrm{\scriptsize tot}}$=1), while the data pointed to a negatively curved spatial section $\Omega_{\textrm{\scriptsize tot}}<1$.
In \cite{aurich3, aurich4} the CMB was investigated for a small compact hyperbolic universe model (an orbifold) with $0.3\le \Omega_{\textrm{\scriptsize tot}} \le 0.6$, and for the nearly flat case with $\Omega_{\textrm{\scriptsize tot}} \le 0.95$,
respectively, containing radiation, baryonic, cold dark matter and $\Lambda$. It was shown that the low multipoles are suppressed even for nearly flat, but hyperbolic models with $\Omega_{\textrm{\scriptsize tot}} \le 0.9$.
For even larger values of $\Omega_{\textrm{\scriptsize tot}} \approx 0.95$, fluctuations of the low multipole moments $C_{l}$ occur, which are typical in the case of a finite volume of the Universe. 
In \cite{aurich5, aurich6} the first-year WMAP data and the magnitude-redshift relation of Supernovae of type Ia have been analyzed in the framework of quintessence models 
and it has been shown that the data are consistent with a nearly flat hyperbolic geometry of the Universe if the optical depth $\tau$ to the surface of last scattering is not too big. 

Furthermore it has been shown \cite{aurich7, aurichsteinerthen, aurich1} that the hyperbolic space form of the Picard universe model, defined by the Picard group which has an infinitely long horn but finite volume, leads e.g.
for $\Omega_{\textrm{\scriptsize matter}}=0.30$ and $\Omega_{\Lambda}=0.65$, to a very small quadrupole and displays very small correlations at angles $\vartheta \ge 60 ^{\circ}$. Even at small angles, $\vartheta \approx10^{\circ}$, $C(\vartheta)$ agrees with the observations much better than the $\Lambda$CDM model.

Depending on certain priors, the WMAP team reported in 2003 from the first-year data \cite{spergel1} for the total energy density $\Omega_{\textrm{\scriptsize tot}} =1.02 \pm 0.02$ together with 
$\Omega_{\textrm{\scriptsize baryon}} =0.044 \pm 0.004$, $\Omega_{\textrm{\scriptsize matter}}=0.27\pm 0.04$, and $h=0.71^{+0.04}_{-0.03}$ for the present-day reduced Hubble constant $h=H_{0}/(100$km~s$^{-1}$~Mpc$^{-1})$ (the errors give the $1\sigma$-deviation uncertainties).
Taken at face value, these parameters hint at a positively curved Universe. Luminet et al. \cite{luminet1} studied the Poincar\'e dodecahedral space which is one of the well-known space forms
with constant positive curvature. In \cite{luminet1} only the first three modes of the Laplacian have been used (comprising in total $59$ eigenfunctions), which in turn restricted the discussion
to the multipoles $l \le 4$. Normalizing the angular power spectrum at $l =4$, they found for $\Omega_{\textrm{\scriptsize tot}}=1.013$, a strong suppression of the quadrupole and a weak suppression of the octopole.
The 2--pcf $C(\vartheta)$ could not be calculated. 

A thorough discussion of the CMB anisotropy and of $C(\vartheta)$ for the dodecahedral topology was carried out in \cite{aurich8} based on the first $10\,521$ eigenfunctions.  
An exact analytical expression was derived for the mean value of the multipole moments $C_{l}$ ($l \ge 2$) for the ordinary Sachs-Wolfe contribution 
(i.e. without the integrated Sachs-Wolfe effect and the Doppler contribution), which explicitly shows that the lowest multipoles are suppressed due to the discrete spectrum
of the vibrational modes. The discrete eigenvalues for all spherical spaces are in appropriate units given by $E_{\beta}=\beta^{2}-1$, where the dimensionless wave
numbers $\beta$ run through a subset of the natural numbers. (Only in the case of the simply connected sphere ${\mathcal S^3}$, $\beta$ runs through all natural numbers.)
In the case of the dodecahedral space there exist no 
even wave numbers, and the odd wave numbers have large gaps since, e.g., the allowed $\beta$-values up to $41$ are given by $\{1, 13, 25, 31, 33, 37, 41\}$, where $\beta=1$, 
corresponding to the zero mode $E_{1}=0$, is subtracted since it gives the monopole. Thus, the spectrum is not only discrete 
but has in addition large gaps (`missing modes') which lead to an additional suppression. The analytical expression for the $C_{l}$'s also leads to an analytical 
expression for the correlation function (due to the ordinary Sachs-Wolfe contribution) which shows the suppression at large scales \cite{aurich8}. The remaining
contributions from the integrated Sachs-Wolfe and Doppler effect were computed numerically. A detailed analysis of the CMB anisotropy for all spherical spaces
was carried out in \cite{aurich2}, and it was shown that only three spaces out of the infinitely many homogeneous spherical spaces are in agreement with the first-year WMAP data. 

The question of the strange alignment of the quadrupole with the octopole, and the extreme planarity or the extreme sphericity of some multipoles has been investigated in
\cite{aurich9} with respect to the maximal angular moment dispersion and the Maxwellian multipole vectors for five multiply connected spaces: the Picard topology in hyperbolic space  
\cite{aurich7, aurichsteinerthen, aurich1}, three spherical spaces (Poincar\'e dodecahedron \cite{luminet1, aurich8}, binary tetrahedron and binary octahedron \cite{aurich2}) and the cubic torus\cite{aurich10}. 
Although these spaces are able to produce the large-scale suppression of the CMB anisotropy, they do not describe the CMB alignment. From the models
considered, the Picard space form reveals the strongest alignment properties.

Already the $3-$year data of WMAP provided a hint that our Universe might be spatially flat \cite{spergel2}. The 2018 results reported by the Planck team \cite{planck7},
combining Planck temperature and polarization data and BAO (baryon acoustic oscillation) measurements, give for the curvature parameter $\Omega_{K}:=1-\Omega_{\textrm{\scriptsize tot}}$ the small value      
$\Omega_{K}=0.0007\pm0.0019$, suggesting flatness to a $1\sigma$ accuracy of $0.2\%$.
Recently, however, a different interpretation has been presented claiming that the data show a preference for a positively curved Universe, noted also in \cite{planck7}
(for references see \cite{efstathiou1}). This problem has been revisited in \cite{efstathiou1}, and when combining with other astrophysical data, it is concluded that
spatial flatness holds to extremely high precision with $\Omega_{K}=0.0004\pm0.0018$ in agreement with Planck \cite{planck7}.
But, also recently, it has been pointed out \cite{Valentino, Valentino2, Handley, Vagnozzi1, Vagnozzi2} that there are inconsistencies between cosmological datasets arising when the FLRW curvature parameter $\Omega_K$ is determined from the data rather 
than constrained to be zero \textit{a priori}. Relaxing this prior also increases the already substantial discrepancy between the Hubble parameter as determined by Planck and local observations to the level of $5\sigma$. 
These different outcomes originate from the comparison of data at the CMB epoch and data from the present-day Universe providing `tensions' for the $\Lambda$CDM model \cite{tensions1}, \cite{tensions2}, \cite{tensions3}. 
Resolving these tensions appears to need a fully general-relativistic description of the curvature evolution \cite{curvaturecrisis}.

Assuming that the spatial section of our Universe is well-approximated by a flat manifold that is furthermore simply connected, it follows that its topology is given by the
infinite Euclidean $3-$space $\mathbb{E}^{3}$. This is exactly the assumption made in the $\Lambda$CDM model which leads to the intriguing discrepancies in the range $\Upsilon$ of large angular scales as discussed above (see figure \ref{fig1}). 
It has been shown in \cite{aurich10,aurich11,aurich12,aurich14} that the simplest spatially flat finite-volume manifold with non-trivial topology i.e. the multiply-connected 
cubic $3-$torus ${\mathcal T}^{3}$ with side length $L$ having the finite volume $L^{3}$, leads in a natural way, without additional assumptions, to the observed suppression at large scales if only
the volume is not too large. For the many previous works on a toroidal universe model, see the references in \cite{aurich10}. 

A modified correlation function, the spatial correlation function, was suggested in \cite{roukema3}, which takes the assumed underlying topology of the dodecahedron into account 
and provides estimates for the orientation of the manifold. This method was applied to the $3-$torus topology in \cite{aurich12}. Another example of topology is provided by the flat slab space \cite{bernui1} with one 
compact direction and two infinite directions. 
A further example is provided by the compact Hantzsche-Wendt manifold for which the ensemble averages of statistical quantities such as the 2--pcf depend on the position of the observer
in the manifold, which is not the case for the $3-$torus topology ${\mathcal T^{3}}$. The suppression of correlations of the 2--pcf is studied in \cite{aurich13}. 
For this topology, the `matched circles-in-the-sky' signature is much more difficult to detect because there are much fewer back-to-back circles compared to the ${\mathcal T^{3}}$ topology.

While the infinite $\Lambda$CDM model is homogeneous and isotropic, the multiply connected torus Universe ${\mathcal T^{3}}$ is still homogeneous and locally isotropic, but no more globally isotropic.
In a flat Universe having three infinite spatial directions such as for the $\Lambda$CDM model, the spectrum of the vibrational modes (i.e. the eigenvalues and eigenfunctions) of the Laplacian is continuous. 
In the case of the $3-$torus topology ${\mathcal T}^{3}$, the CMB temperature anisotropies
$\delta T$ over the $2-$sphere ${\cal S}^{2}$ are calculated by using
the vibrational modes of the Laplacian with periodic conditions
imposed by the cubic fundamental domain without boundary\cite{aurich10}.
The discrete eigenvalues of the Laplacian are then given by
\begin{equation}
\label{equ:eigenlaplacian}
{E_{\boldsymbol{n}}}=\left(\frac{2\pi}{L}\right)^{2} \boldsymbol{n}^{2} ~~~ \nonumber{\textrm{with}} ~ \boldsymbol{n}=(n_{1},n_{2},n_{3}) \in \mathbb{Z}^{3} \;\;.
\end{equation}
Thus, the wave number spectrum of ${\mathcal T}^{3}$ is discrete and countably infinite consisting of the distinct wave numbers
\begin{equation}
\label{equ:km}
{k_{m}}=\frac{2\pi}{L} \sqrt{m}\ , \ m=0, 1, 2, ...\;,
\end{equation}
i.e. there is no ultraviolet cut-off at large wave numbers. There are gaps between consecutive wave numbers,
\begin{equation}
\label{equ:kmplus1}
{k_{m+1}-k_{m}}= \frac{\pi}{L}\frac{1}{\sqrt{m}}\left[ 1-\frac{1}{4m} + {\mathcal O} \left(\frac{1}{m^2}\right) \right] \ , \ m \rightarrow \infty\ ,  
\end{equation}
which tend to zero asymptotically. However, the wave numbers are degenerate, i.e. they possess multiplicities $r_{3}(m)$, where $r_{3}(m)$ is a very irregular, number-theoretical function 
with increasing mean value, which counts the number of representations of $m \in \mathbb{N}_{0}$ as a sum of $3$ squares of integers, where representations with different orders and different signs are counted as distinct.
For example, $r_{3}(0)=1$, $r_{3}(1)=6$, $r_{3}(2)=12$, $r_{3}(3)=8$, $r_{3}(4)=6$, $r_{3}(5)=24$. ($r_{3}(m)$ has been already studied by Gauss.)
Weyl's law provides the asymptotic growth of the number $N(K)$ of all vibrational modes in the $3-$torus with 
$|\boldsymbol{k_{\boldsymbol{n}}}|:=\sqrt{E_{\boldsymbol{n}}} \le~K$,
\begin{equation}
\label{equ:Nk}
{N(K)}\sim\frac{V}{6\pi^{2}}K^{3} \; , \ K \rightarrow \infty\ ,
\end{equation}
where $V=L^{3}$ is the volume of the torus manifold (see for instance \cite{aurich4,aurich8,aurich2} and the review \cite{arendt1}).
For example, in \cite{aurich14}, the first $50\,000$ distinct wave numbers were taken into account comprising in total $61\,556\,892$ vibrational modes which allowed to compute the multipoles up to $l = 1\,000$. 
There is, however, in the case of the CMB anisotropy in a torus universe model, a cut-off at small wave numbers, i.e. an infrared cut-off, $|\boldsymbol{k_{\boldsymbol{n}}}|\ge2\pi/L=k_{1}$, 
since the zero mode $|\boldsymbol{k_{0}}|=0$ has been subtracted, as was first pointed out by Infeld in the late forties \cite{infeld1}.  

In this paper, the cosmological lengths are expressed in terms of the Hubble length denoted $L_H=c/H_{0}$ as in \cite{aurich10, aurich12}.  
The value of the reduced Hubble constant today according to Planck 2015 \cite{planck3} was $h=(0.6727 \pm0.0066)$ ($68\%$ limits) and is used in the tables \ref{table1} 
and \ref{table2} of this study, giving a Hubble length of $L_H=(4.4453^{+0.0386}_{-0.0379})$ Gpc. The value determined from the most recent analysis of Planck from the $\Lambda$CDM model 
in 2019 \cite{efstathiou2} is very close, i.e. at $h=(0.6744\pm0.0058)$ ($68\%$ limits).
The $3-$torus and $\Lambda$CDM simulations presented in this work are calculated using the Planck 2015 cosmological parameters. But, given the small $N_{\textrm{\scriptsize side}}=128$ and $l_{\textrm{\scriptsize max}}=256$ and 
the strong Gaussian smoothing scale of 2$^{\circ}$ f.w.h.m., suppressing the sharp CMB structures at the first acoustic peak and beyond, differences between using the Planck 2015 or the 
Planck 2019 cosmological parameters to generate the CMB temperature maps are not expected in terms of cosmic topology. 
It is only when considering the improved polarization data of Planck legacy 2018 that differences might be expected for cosmic topology.

For the CMB in a universe model with $3-$torus topology and with an optimally determined torus side length of $L\approx 3.69 L_H$, the 2--pcf is nearly vanishing for 
large angles \cite{aurich10, aurich12}, fitting much better to the 2--pcf of the observed maps than those of the $\Lambda$CDM model. In the case of the slab space manifold (only 
one compact direction \cite{bernui1}) the match with the Planck 2015 CMB maps 2--pcf is good, once the slab is optimally oriented with respect to our galactic plane and for 
an optimal slab thickness close to $4.4L_H$ (for the same $H_0$ of Planck 2015). Also good is the 2--pcf match for any angle separation \cite{bernui1}, except for the 
angles beyond $150^{\circ}$ where the remnants of galactic foreground pollution in the Planck maps could explain the non-zero and negative value of the correlation at the largest scales. 

Another signature of multiply connected topology, the `matched circles-in-the-sky' (thereafter CITS) was and is much tested on the COBE, WMAP and Planck CMB temperature and polarization maps
(see \cite{planck4}, \cite{cornish1 , cornish2, roukema1, roukema2}, \cite{aurich10} and \cite{cornish3, vaudrevange1, aurich15, gomero1}).  
The CITS signal is based on the fact that the metric perturbation at the surface of last scattering (SLS) is responsible for a large part of the CMB signal,
since the metric perturbation is mapped by the $3-$torus group to the identified points on the SLS. Other contributions to the CMB signal deteriorate the CITS signal, such as
the Doppler contribution which is projected from different lines of sight. Another deteriorating effect is due to the integrated Sachs-Wolfe (ISW) effect, which describes 
the changes along the photon path from the SLS to the observer, where again the points along the path are not identified due to the topology. There are hints that this contribution 
is larger than expected from the $\Lambda$CDM model, so that the CITS signal might be less pronounced than derived from $\Lambda$CDM simulations.
The late time ISW effect due to the supervoids is stronger than expected from the $\Lambda$CDM model measured by $A_{\hbox{\scriptsize ISW}} = \Delta T^{\hbox{\scriptsize data}}/\Delta T^{\hbox{\scriptsize theory}}$ \cite{kovacs1} which should be close to $1$. 
The Dark Energy Survey (DES) collaboration finds an excess amplitude $A_{\hbox{\scriptsize ISW}}$=4.1 $\pm$2.0 \cite{kovacs1} and when they combine their data with the independent 
Baryon acoustic Oscillations Spectroscopic Survey (BOSS) data, even an excess ISW signal of supervoids with $A_{\hbox{\scriptsize ISW}}$=5.2 $\pm$1.6 is revealed.
For a summary, see figure 5 in \cite{kovacs1} and references in this publication.
It thus seems to be premature to exclude non-trivial topologies due to the non-observations of the CITS signal on the SLS. 

The other tested signature of a $3-$torus multiply connected topology, the covariance matrix, entails no conclusive results, e.g. \cite{aurich10}. 
It is then of great importance to confirm the possibly multiply connected nature of our Universe suggested by the vanishing 2--pcf through complementary methods using different observables and implemented with other morphological or topological descriptors. 

In the present study we consider a global scalar which appears to provide a complementary method of detecting a multiply connected Universe from the CMB map analysis. 

\newpage

This paper is organized as follows.
In section \ref{sec:definition}, the conventions and definitions of the quantities used are given, and the normalized standard deviation $\rho$ of the CMB
temperature gradient field, the central object of this investigation, is introduced.
Section \ref{sec:dependence} presents the main outcome: the hierarchical dependence between the size of the topological fundamental cell of the universe model and the normalized 
standard deviation $\rho$ of the temperature gradient.
This result is based on the analysis of five ensembles of cubic $3-$tori ${\mathcal T}^3$ of increasing size, as well as one ensemble of the infinite $\Lambda$CDM model,
and the Planck CMB maps. 
In section \ref{sec:comparison} two CMB maps in the cubic $3-$torus topology at small ($L=0.5L_H$) and large ($L=3.0L_H$) side lengths illustrate how much the different spectra of vibrational modes 
influence the scale of spatial features on the CMB map. The average 2--pcfs of those different $3-$torus sizes are shown and commented. 
Section \ref{sec:torusisotropy} is dedicated to quantify the level of isotropy and homogeneity of the $3-$torus CMB maps.
In section \ref{sec:discussion} the ingredients of the Boltzmann physics used in simulations are presented. 
We develop on the attempts and limitations to predict the relation between $L$ and $\rho$.
In section \ref{sec:conclusion} we conclude that $\rho$ can serve as a sensitive probe and leads to a complementary test for a multiply connected Universe, although the preferred 
side length is found to be mildly smaller than the value derived from the 2--pcf analysis. Finally, we show and discuss the fact that results from the $\rho$-analysis 
would augment the list of the CMB anomalies.

\section{The normalized standard deviation $\rho$ of the temperature gradient field $\boldsymbol{G}$}
\label{sec:definition}
The CMB temperature fluctuation $\delta T (\boldsymbol{\hat n}):=T(\boldsymbol{\hat n})-T_{0}$ is defined as the
difference between the direction-dependent temperature $T(\boldsymbol{\hat n})$ and
the monopole $T_{0}:=T_{\hbox{\scriptsize CMB}}$, with $T_{\hbox{\scriptsize CMB}}=(2.7255 \pm 0.0006K)$ \cite{fixsen1,planck3}.
On the unit sphere ${\mathcal S}^2$, we write the metric in spherical coordinates $(\vartheta, \varphi)$, 
\begin{equation}
\label{equ:S2metric}
{{\mathrm d}s^{2}}:={\mathrm d}\vartheta^{2}+\sin^{2}\vartheta \, {\mathrm d}\varphi^{2}\;,
\end{equation}
and denote the unit vector by
$\boldsymbol{\hat n}=\boldsymbol{\hat n}(\vartheta, \varphi)$.
The angular average of $\delta T(\boldsymbol{\hat n})$ vanishes,
\begin{equation}
\label{equ:mu}
\frac{1}{4\pi} \int_{{\cal S}^2} {\mathrm d}^2 \boldsymbol{\hat n} \ \delta T(\boldsymbol{\hat n})  = 0 \ .
\end{equation}
Averaging also over the possible positions from which the CMB is observed, one obtains:
\begin{equation}
\int_{{\cal S}^2} {\mathrm d}^2 \boldsymbol{\hat n} \ \mu (\boldsymbol{\hat n}) = 0 \ ,
\end{equation}
with $\mu (\boldsymbol{\hat n}): = \langle \delta T(\boldsymbol{\hat n})\rangle$.
Here, the brackets denote an ensemble average at fixed $\boldsymbol{\hat n}$. Similarly, we define
the ensemble average of the variance of $\delta T(\boldsymbol{\hat n})$,
\begin{equation}
\label{equ:sigma0sq}
\sigma_{0}^{2} (\boldsymbol{\hat n}):=\left<[\delta T(\boldsymbol{\hat n}) - \mu (\boldsymbol{\hat n})]^{2}\right>\ .
\end{equation}
Assuming that the Universe is homogeneous and isotropic \textit{on average}, all averages
$\langle \delta T(\boldsymbol{\hat n})\delta T(\boldsymbol{\hat n}')\delta T(\boldsymbol{\hat n}'')\cdots\rangle$ are 
rotationally invariant functions of $\boldsymbol{\hat n}, \boldsymbol{\hat n}', \boldsymbol{\hat n}'', \cdots$, and thus
$\mu$ and $\sigma_0^2$ are independent of $\boldsymbol{\hat n}$. In this case it follows that $\mu = 0$ \footnote{For an ensemble of $n$ maps we have
$
{\langle\mu\rangle}:= (1/n) \sum_{j=1}^{n}\mu_{j}.
$
After subtraction of the monopole and the dipole (this is done for all the maps studied in this work), 
we can verify that over the $100\, 000$ simulation maps of the $\Lambda$CDM and the $3-$torus models, $\vert\langle\mu\rangle\vert$ is numerically extremely small ${\mathcal O}(10^{-7}\mu K)$ 
without mask and ${\mathcal O}(10^{-4}\mu K) - {\mathcal O}(10^{-3}\mu K)$ after mask pixel suppression.} and 
\begin{equation}
\sigma_0^2 = \langle [ \delta T(\boldsymbol{\hat n})]^2 \rangle \ .
\end{equation}
Since the correlation function \eref{equ:2--pcf} is a function of $\boldsymbol{\hat n}\cdot\boldsymbol{\hat n}' = \cos{\vartheta}$, it can be expanded in Legendre polynomials,
\begin{eqnarray}
\label{Cobs}
C^{\rm obs}(\vartheta) &=& \frac{1}{8\pi^2} \int_{{\cal S}^2} {\mathrm d}^2 \boldsymbol{\hat n}_1 \int_{{\cal S}^2}  {\mathrm d}^2 \boldsymbol{\hat n}_2 \ \delta \left(\boldsymbol{\hat n}_1 \cdot\boldsymbol{\hat n}_2 - \cos{\vartheta}\right)\delta T(\boldsymbol{\hat n}_1)\delta T(\boldsymbol{\hat n}_2) \nonumber\\ &=& \frac{1}{4\pi} \sum_{l =1}^{\infty} (2l+1) C_{l}^{\rm obs} \; P_l (\cos\vartheta) \ , 
\end{eqnarray}
with the multipole moments 
\begin{equation}
\label{alm}
\fl
C_{l}^{\rm obs}: = \frac{1}{4\pi}  \int_{{\cal S}^2}  {\mathrm d}^2 \boldsymbol{\hat n} \int_{{\cal S}^2} {\mathrm d}^2 \boldsymbol{\hat n}' \
P_l (\boldsymbol{\hat n}\cdot \boldsymbol{\hat n}') \ \delta T(\boldsymbol{\hat n})\delta T(\boldsymbol{\hat n}')
= \frac{1}{2l+1} \sum_{m= -l}^{l} \vert a_{lm}\vert^2 \ ,
\end{equation}
and where the complex coefficients $\lbrace a_{lm}\rbrace$ are the coefficients of the expansion of $\delta T(\boldsymbol{\hat n})$ into spherical harmonics, $Y_l^{\;m}(\boldsymbol{\hat n})$, on the full sky. The observed angular power spectrum is then given by
\begin{equation}
\left( \delta T_l^{\rm obs} \right)^2 := \frac{l(l+1)}{2\pi} C_l^{\rm obs} \ .
\end{equation} 
Note that equations \eref{Cobs} and \eref{alm} hold without any theoretical assumptions on $\delta T(\boldsymbol{\hat n})$
(provided the integrals and series converge).

Assuming that the Universe is homogeneous and isotropic on average, the ensemble average of the full-sky correlation function is rotationally invariant and satisfies
\begin{equation}
\label{Ctheta}
C(\vartheta) : = \langle \delta T(\boldsymbol{\hat n})\delta T(\boldsymbol{\hat n}') \rangle 
= \frac{1}{4\pi} \sum_{l=1}^{\infty} (2l+1) C_l \; P_l (\boldsymbol{\hat n}\cdot \boldsymbol{\hat n}') \ ,
\end{equation}
with the multipole moments 
\begin{equation}
\label{alm2}
C_l : = \langle \vert a_{lm}\vert^2 \rangle 
\end{equation}
(independent of $m$). From \eref{alm} and \eref{alm2} follows that 
\begin{equation}
\label{alm3}
\langle C_l^{\rm obs} \rangle = C_l \ .
\end{equation}
From \eref{alm2} and \eref{alm3} one finds the normalized variance of $C_l - C_l^{\rm obs}$, i.e. 
the cosmic variance:
\begin{equation}
\label{cosmicvariance}
\left\langle \left(   \frac{C_l - C_l^{\rm obs}}{C_l} \right)^2 \right\rangle = - 1+ \frac{1}{(2l+1)^2 C_l^2} 
\sum_{m = -l}^{l} \sum_{m' = -l}^{l} \langle \vert a_{lm}\vert^2 \vert a_{lm'}\vert^2 \rangle \ .
\end{equation}
If we furthermore assume that $ \delta T({\boldsymbol{\hat n}})$ is a Gaussian random field on ${\cal S}^2$, it follows that the $\lbrace a_{lm}\rbrace$ are complex Gaussian random variables which, however, does not imply that
also the $C_l$'s are Gaussian random variables. The cosmic variance \eref{cosmicvariance} simplifies in the Gaussian case and is given by
\begin{equation}
\left\langle \left(   \frac{C_l - C_l^{\rm obs}}{C_l} \right)^2 \right\rangle = \frac{2}{2l+1} \ .
\end{equation}
(For the case with mask, the reader is directed to equations (22)-(24) in \cite{aurich13}.)

Over the $2-$sphere support of the CMB temperature anisotropy map we also define $\boldsymbol{G}$, the gradient field, dependent on the spherical coordinates $\vartheta$ and $\varphi$. In terms of its components,
\begin{equation}
\label{equ:gtheta}
G_{\vartheta}:=\frac{\partial \delta T}{\partial \vartheta}\;,
\end{equation}
and
\begin{equation}
\label{equ:gphi}
G_{\varphi}:=\frac{1}{\sin \vartheta}\frac{\partial \delta T}{\partial \varphi}\;.
\end{equation}
The variance $\sigma_{1}^{2}$ of the local temperature gradient
is defined by an average over the directions,
\begin{equation}
\label{equ:sigma1sq1}
{\sigma_{1}^{2}}:=\left<\nabla_{1} \delta T ({\boldsymbol{\hat n}})\nabla^{1} \delta T ({\boldsymbol{\hat n}}) + \nabla_{2} \delta T ({\boldsymbol{\hat n}})\nabla^{2} \delta T ({\boldsymbol{\hat n}})\right>\;,
\end{equation}
where in spherical coordinates the covariant derivatives are given by
\begin{equation}
\label{equ:sigma1sq1theta}
\nabla_{1} \delta T ({\boldsymbol{\hat n}})\nabla^{1} \delta T ({\boldsymbol{\hat n}}) = \delta T_{,\vartheta}\delta T^{,\vartheta} = G_{\vartheta}^2 = \left(\frac{\partial \delta T}{\partial \vartheta}\right)^{2}\;, 
\end{equation}
and 
\begin{equation}
\label{equ:sigma1sq1phi}
\nabla_{2} \delta T ({\boldsymbol{\hat n}})\nabla^{2} \delta T ({\boldsymbol{\hat n}})=\delta T_{,\varphi}\delta T^{,\varphi} = G_{\varphi}^2 = \left(\frac{1}{\sin \vartheta}\frac{\partial \delta T}{\partial \varphi}\right)^{2}\;.
\end{equation}
If the CMB sky map is an isotropic and homogeneous Gaussian random field having a negligible mean $\mu$ (hereafter IHG properties, IHG standing for isotropic, homogeneous and Gaussian of zero mean),  
the ensemble average of the CMB is statistically determined by its 2--pcf $C(\vartheta)$, equations \eref{Ctheta} and \eref{alm2}.
Under this condition the components $G_{\vartheta}$ and $G_{\varphi}$ of the gradient vector $\boldsymbol{G}$, equations \eref{equ:gtheta} and \eref{equ:gphi}, are Gaussian random variables with zero mean and identical variance $\sigma_{1}^{2}/2$.

\smallskip

The field of CMB temperature anisotropies $\delta T(\boldsymbol{\hat n})$ is discretized into pixels of the HEALPix tessellation 
$\delta T_{i}:=\delta T(\boldsymbol{\hat n}_{i})$, and for the purpose of this investigation, $\sigma_{1}^{2}$ is calculated in pixel space in spherical coordinates (see also the formulas in the non-discretized case, 
equations (29) and (30) in \cite{monteserin1}) as the average\footnote{This is implemented using a modified version of the HEALPix Fortran subroutine `alm2map$\_$der' and its function `der1'. } expanded into
\begin{equation}
\label{equ:sigma1sq3}
{\sigma_{1}^{2}}:= \left< G_{\vartheta}^2+G_{\varphi}^2\right> = \frac{ \sum_{i=0}^{\hbox{\scriptsize npixels}-1} \left[ \left(\frac{\partial \delta T_{i}}{\partial \vartheta}\right)^{2} + \left(\frac{1}{\sin \vartheta}\frac{\partial \delta T_{i}}{\partial \varphi}\right)^{2} \right] }{\hbox{npixels}}\;.
\end{equation}
The reader may refer to mathematical definitions, developments and discussions related to scalar statistics on the CMB spherical support manifold in \cite{bond1}, \cite{schmalzing1}, \cite{monteserin1}, \cite{aurich14}. 
Under the assumption that $\delta T(\boldsymbol{\hat n})$ is an isotropic and homogeneous random field on average, $\sigma_{1}^{2}$ can be calculated in the spherical harmonic space as
\begin{equation}
\label{equ:sigma1sq5}
{\sigma_{1}^{2}}:= \sum_{l=l_{\hbox{\scriptsize min}}}^{l_{\hbox{\scriptsize max}}}C_{l}\frac{l(l+1)(2l+1)}{4\pi}\;,
\end{equation}
where $C_{l}$ are the multipole moments \eref{alm2}, monopole and dipole are subtracted (i.e. $l_{\hbox{\scriptsize min}}=2$) and $l_{\hbox{\scriptsize max}}=256$. Under the same isotropy condition, the variance of
$\delta T(\boldsymbol{\hat n})$ reads:
\begin{equation}
\label{equ:sigma0sq5}
{\sigma_{0}^{2}}:= \sum_{l=l_{\hbox{\scriptsize min}}}^{l_{\hbox{\scriptsize max}}}C_{l}\frac{2l+1}{4\pi}\;.
\end{equation} 
If the CMB sky maps possess the IHG properties, they are statistically completely determined by the multipoles $C_l$. 
Also, the equivalence between the 2--pcf $C(\vartheta)$ and the power spectrum ($\delta T_{l}^{2}:=l(l+1)C_{l}/2\pi$) only holds if the CMB over the whole $2-$sphere is observable.

\smallskip

We define $\rho$ as the normalized standard deviation of the gradient field $\boldsymbol{G}$ of temperature anisotropy over a single map,
\begin{equation}
\label{equ:rho}
{\rho}:= \sqrt{ \frac{\left<G_{\vartheta}^2+G_{\varphi}^2\right>}{\sigma_{0}^{2}} } = \sqrt{\frac{\sigma_{1}^{2}}{\sigma_{0}^{2}}}\;\;,
\end{equation}
while the mean in terms of $\rho$ for an ensemble of $n$ maps is given by
\begin{equation}
\label{equ:rho-aver}
{\langle\rho\rangle}:= \frac{\sum_{j=1}^{n}\rho_{j}}{n}\;\;.
\end{equation}
While searching for possible non-Gaussianities in the CMB maps using Minkowski functionals \cite{buchert1}, one of us (FS) proposed in 2012 the normalized variance of the CMB gradient as a new signature of a multiply connected nature of the Universe. 
First applications to cubic tori of different volumes indeed revealed \cite{lustig1} 
that there is a hierarchical dependence of $\rho$ as a function of $L$ the side length of the torus.
Note that the ratios $\sigma_{1} / \sigma_{0}$ and respectively $\sigma_{1}^{2} / \sigma_{0}^{2}$, appear in the definition of the Gaussian prediction of the second Minkowski functional (MF) and respectively the third 
MF of a random field on the $2-$sphere ${\mathcal S}^2$.
For comprehensive definitions of 
random fields and Minkowski functionals of excursion sets, see \cite{adler1}, \cite{tomita1, tomita2}, \cite{bond1}, \cite{schmalzingbuchert, schmalzing1}, \cite{aurich14}, \cite{buchert1}. 

Obviously, $\rho$ defined as a ratio does not depend on an overall normalization constant of the temperature field.
While a comparison of maps using only $\sigma_{0}$ or $\sigma_{1}$ or the 2--pcf requires the normalization of the temperature anisotropy field.
We shall develop on this application of normalization for our ensembles of $3-$torus maps in sections \ref{sec:comparison} and \ref{sec:discussion}.

\bigskip
\begin{figure}[!htb]
\includegraphics[height=0.35\textheight,width=1.00\textwidth, angle=0]{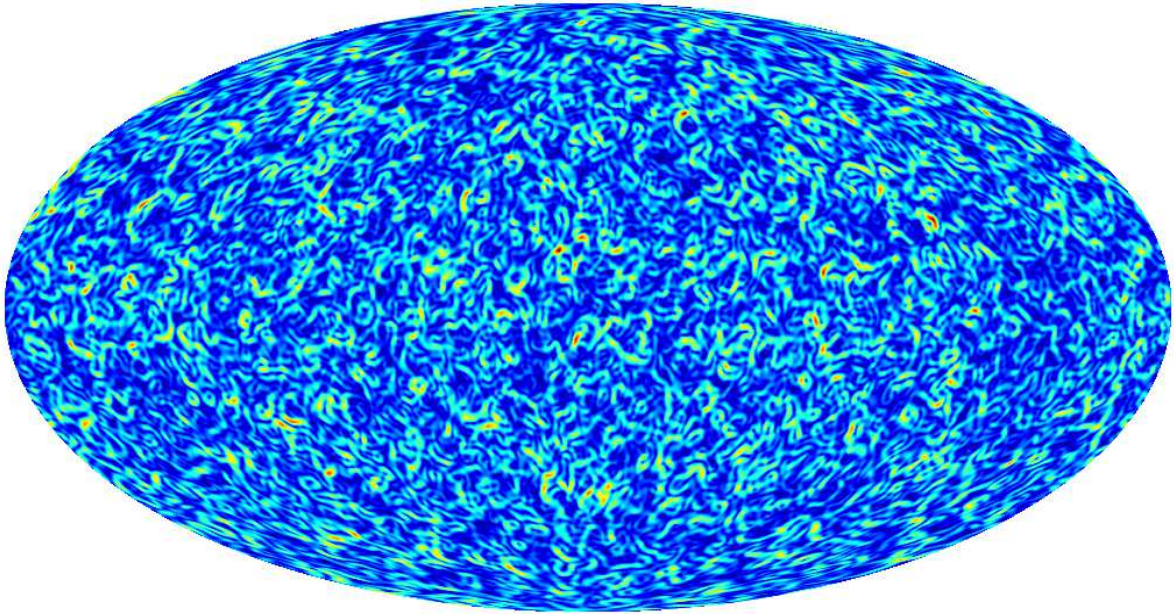}
\caption{The map of the normalized norm of the temperature gradient field $\boldsymbol{G}$, as defined in 
equation \eref{equ:rho-i} and calculated from a CMB map of the $3-$torus at $L=1.0L_H$ at a resolution of $N_{\hbox{\scriptsize side}} =128$, $l_{\hbox{\scriptsize range}}=[2,256]$ and $\vartheta_{G} =2^{\circ}$. In the online version (respectively in the offprint version), the strongest local gradients 
appear in red (dark) and the weakest gradients in dark-blue (black). While the most interesting features, the numerous iso-contour patterns, are shown in light-blue (white).}
\label{fig2}
\end{figure}
In order to provide an illustration of the quantity from which $\rho$ is derived by averaging $\boldsymbol{G}$,
figure \ref{fig2} shows in Mollweide projection the map of
\begin{equation}
\label{equ:rho-i}
{\rho_{i}}:= \sqrt{ \frac{G_{\vartheta,i}^2+G_{\varphi,i}^2}{\sigma_{0}^{2}} } = \sqrt{\frac{\sigma_{1,i}^{2}}{\sigma_{0}^{2}}}\;\;,
\end{equation}
where $i$ denotes a pixel index, for one CMB map of the $3-$torus simulations at a side length of $L=1.0L_H$.
The resolution parameters are the ones applied to all the maps all along the present study i.e. $N_{\hbox{\scriptsize side}} =128$, $l_{\hbox{\scriptsize range}}=[2,256]$ and a Gaussian smoothing $\vartheta_{G} =2^{\circ}$ f.w.h.m. The Gaussian smoothing is defined by $C_{l} \rightarrow C_{l}|F_{l}|^{2}$ with
\begin{equation}
\label{equ:gausskernel}
{F_l}=\exp{\left( -\frac{\alpha^2\vartheta_{G}^2}{2} \; l(l+1) \right)}\;
\end{equation}
and $\alpha=\pi/(180 \sqrt{8 \ln{2}})$, which is obtained in the limit $\alpha \vartheta_G \ll 1$ from the Gaussian kernel on $\mathcal S^{2}$.

\section{A hierarchical dependence of the size of the fundamental cell versus $\rho$}
\label{sec:dependence}

The following analysis is based on five ensembles of the cubic $3-$torus ${\mathcal T}^3$
topology belonging to different sizes of the fundamental cell,
and one ensemble of the infinite $\Lambda$CDM model (with a simply connected topology).
The five $3-$torus ensembles belong to the side lengths
$L/L_H=0.5, 1.0, 1.5, 2.0 \hbox{ and } 3.0$.
Each ensemble consists of $100\,000$ realizations leading to $100\,000$ CMB sky maps.\footnote{The simulation of the map ensembles for larger side lengths of the torus is computationally expensive, 
typically months for a hundred core cluster.}
In order to generate a realization of the ensemble, a Gaussian random number of unit variance and zero mean is multiplied by each eigenmode belonging to a wavenumber
$\boldsymbol{k_{\boldsymbol{n}}}$, see equation \eref{equ:km}. 
The CMB maps of the $3-$torus and the infinite $\Lambda$CDM model are computed
using the cosmological parameters according to Planck 2015 \cite{planck3}.
The CMB maps are analyzed at a HEALPix resolution of $N_{\hbox{\scriptsize side}}=128$ ($196608$ pixels of diagonal $27.5'$, i.e. a pixel side length of $19.4'$) with $l_{\hbox{\scriptsize max}}=256$, 
and are smoothed with $\vartheta_{G}=2^{\circ}$. 

\begin{figure}[!htb]
\includegraphics[height=0.68\textheight,width=0.68\textwidth, angle=270]{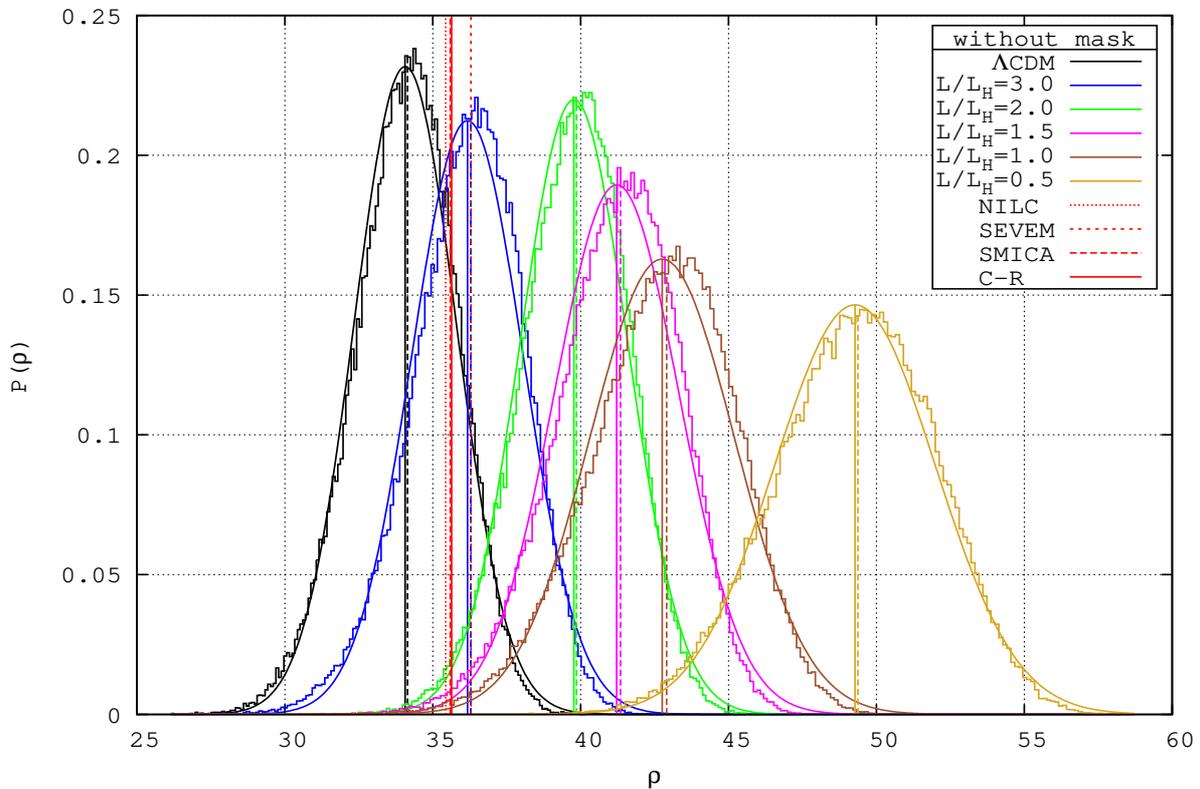}
\caption{The histograms of $\rho$ without foreground mask. Presented in solid lines are the PDF histograms and the Gaussian distributions for the ensembles of $100\,000$ model maps. From right to left: 
in colour for the version online (gray-scales for the offprint version) the $3-$torus at $L=0.5L_H$ in light brown, $L=1.0L_H$ in brown, $L=1.5L_H$ in magenta, $L=2.0L_H$ in green, $L=3.0L_H$ in blue and the $\Lambda$CDM model in black.
The Gaussian PDFs are computed from the means and the variances given in table \ref{table1} and illustrate the deviation of the PDFs from a symmetrical distribution. The means are shown with a vertical solid line and the medians with a vertical dashed line. The mean $\langle\rho\rangle$ for each of the four Planck maps is shown with a vertical line in red for the online version (NILC in small dots, SEVEM in small dashes, SMICA in large dashes and Commander-Ruler as a solid line).} 
\label{fig3}
\end{figure}

\begin{figure}[!htb]
\includegraphics[height=0.68\textheight,width=0.68\textwidth, angle=270]{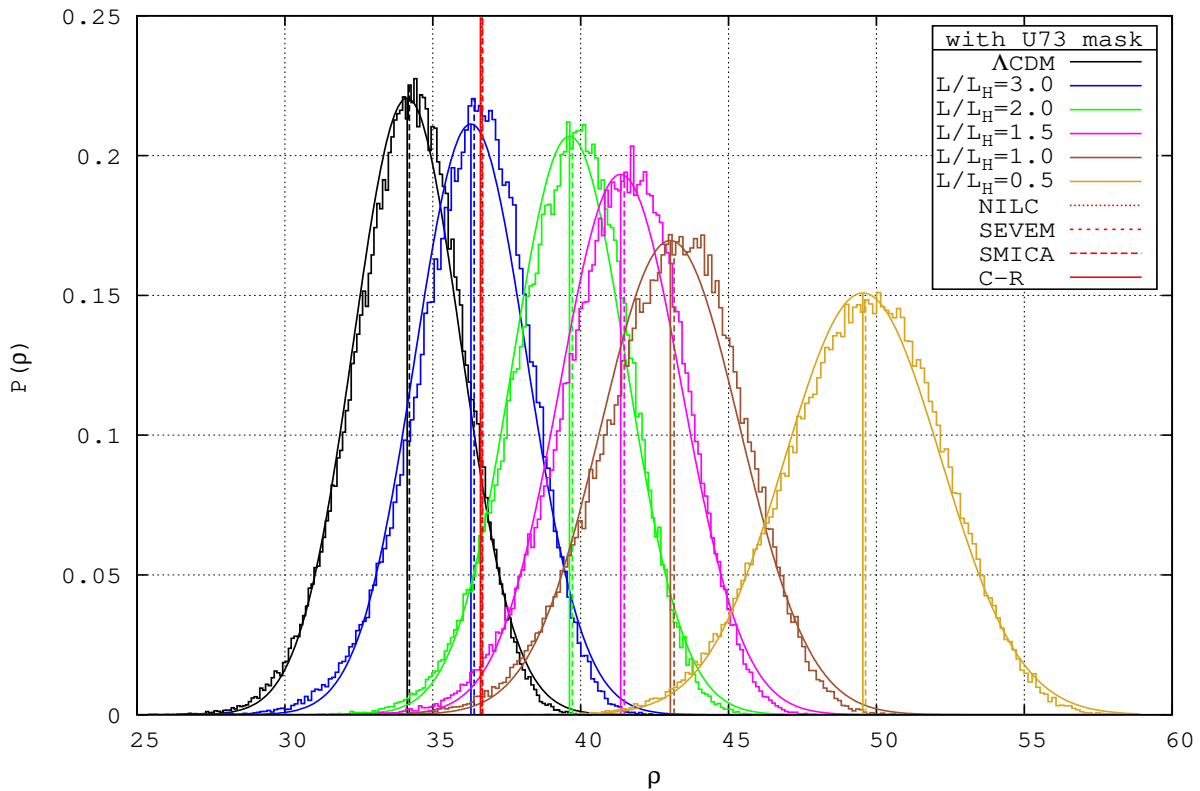}
\caption{Same as figure \ref{fig3} for the histograms of $\rho$, here with exclusion of the U73 mask pixels. 
The Gaussian PDFs are computed from the means and the variances given in table \ref{table2}. 
} 
\label{fig4}
\end{figure}

For each set of $100\,000$ maps, the probability distribution functions (PDFs) of $\rho$ are shown for the five cubic $3-$torus side lengths $L$, and for the infinite $\Lambda$CDM model, as histograms in figure \ref{fig3} (unmasked case) and figure \ref{fig4} (masked case). 
All distributions are unimodal with a pronounced peak. We present in tables \ref{table1} and \ref{table2} the mean value $\langle\rho\rangle$, 
the median $\rho$-value (hereafter denoted median), the standard deviation $\Sigma$, the skewness coefficient
\begin{equation}
\label{equ:gamma1}
{\gamma_{1}}:=\frac{m_{3}}{\Sigma^{3}}\;\;,
\end{equation} 
and the excess kurtosis
\begin{equation}
\label{equ:gamma2}
{\gamma_{2}}:=\frac{m_{4}}{\Sigma^{4}}-3\;\;,
\end{equation}
where $m_{n}$ denotes the $n^{th}$ central moment of a given distribution (see e.g. \cite{buchert1}).

In this paper we do not discuss a theoretical model for the 
PDF $P(\rho)$, which has been studied by two of us (RA and FS) \cite{aurich21}.
In this model, the PDFs 
of the random variables $\sigma_0$ and $\sigma_1$, respectively, are approximated by truncated Gaussian distributions (see ~\ref{appA}). Under this assumption an analytic expression for $P(\rho)$ is derived in \cite{aurich21} 
describing a unimodal skewed distribution that agrees reasonably well with, for example, the histogram of the 
$3-$torus with side length $L=2L_H$, shown in figure~\ref{fig3}. Thus, the model yields a first approximation to $P(\rho)$.
The deviations from the actual histograms is due to the fact that the PDF of $\sigma_0$ possesses a definite non-Gaussian component, whereas the PDF of $\sigma_1$ only shows a small deviation from a Gaussian behaviour. 
The histograms presented in figure~\ref{fig3} are indeed unimodal, but not Gaussian\footnote{The deviation from Gaussianity does not necessarily imply a violation of the IHG properties.}. 
In order to visualize a possible non-Gaussianity of $P(\rho)$, we shall compare in figures \ref{fig3} and \ref{fig4} the histograms with a Gaussian PDF.

Since $\rho$ is by definition a strictly positive random variable, the appropriate Gaussian PDF to compare with is not the standard normal distribution 
defined on the whole line but rather a truncated normal distribution defined only on the positive half-line. 
Thus, the Gaussian PDF to be applied in this situation should \textit{a priori} be a one-sided truncated Gaussian probability distribution function.
For the construction of the truncated Gaussian we refer to \ref{appA}. There it is shown that the deviations of the truncated Gaussian PDF from the standard normal distribution are, however, extremely small in the case considered here. Therefore, we compare the histograms in figures \ref{fig3} and \ref{fig4} with the standard Gaussian PDF
fixed by the mean values $\langle\rho\rangle$ and the variance $\Sigma^{2}$ given in tables \ref{table1} and \ref{table2}.

\smallskip
\noindent
A Gaussian random variable has the following unique characteristic properties:
\begin{itemize}
\item[--] Its PDF maximizes the (differential) entropy among all probable continuous distributions with fixed first and second moment, and in general among all unimodal distributions.

\item[--] All higher odd moments and all cumulants with $n \ge 3$ are identically zero, i.e. in particular $\gamma_{1}$ = $\gamma_{2}$ = 0.

\item[--] Furthermore, one can show (Marcinkiewicz's theorem \cite{marcinkiewicz1}) that the normal distribution is the only distribution having a finite number of non-zero cumulants.

\item[--] It holds the equality `mean' = `median' = `mode' (where `mode' is defined as the location of the maximum of the unimodal PDF). \\
\end{itemize}

\noindent
Thus, $\gamma_{1}$, $\gamma_{2}$ as well as all higher cumulants and the differences 
\begin{equation}
\label{equ:delta1and2}
{\delta_{1}}:=\textrm{median} - \langle\rho\rangle \quad ; \quad 
{\delta_{2}}:=\textrm{mode} - \langle\rho\rangle \ ,
\end{equation}
can serve as indicators of non-Gaussianity of $P(\rho)$. There exists the general bound (Mallows' bound) for all PDFs with $\Sigma < \infty$:
\begin{equation}
\label{equ:absdelta1}
{|\delta_{1}|} \le \Sigma\;\;,
\end{equation}
and for any unimodal PDF there is the sharper bound
\begin{equation}
\label{equ:abs2delta1}
{|\delta_{1}|} \le \sqrt{\frac{3}{5}}~\Sigma \approx 0.775~\Sigma\;\;.
\end{equation}

Tables \ref{table1} and \ref{table2} show that $\delta_{1} > 0$ for all tori, and thus we can consider the normalized ratio $\delta_{1}/\Sigma$ as another measure of non-Gaussianity.  
A possible non-Gaussianity may be considered as small, if $\delta_{1}/\Sigma$ is smaller by a factor of $10$ than the upper bound \eref{equ:abs2delta1}, i.e. if $\delta_{1}/\Sigma \le 0.078$ holds.

Some general properties of these histograms of $\rho$ arise, independently of taking into account the U73 union mask:
\begin{itemize}
\item[--] All PDFs of $\rho$ show a systematically weak negative skewness $\gamma_{1}$ which is true also for the infinite $\Lambda$CDM sample. 
This skewness is less pronounced for the torus at $L=0.5L_H$.

\item[--] The PDFs for the $3-$torus at $L=0.5L_H$ are platykurtic, i.e. with a small negative excess kurtosis $\gamma_{2}=-0.115$ (no mask) and $\gamma_{2}=-0.109$ (U73 mask). 

\item[--] The PDFs of $L=1.0$, $2.0L_H$ and the $\Lambda$CDM are almost mesokurtic with $\gamma_{2}$ very small and positive ($\gamma_{2}\le 0.085$). 

\item[--] The PDFs of $L=1.5$ and $3.0L_H$ are leptokurtic i.e. with $\gamma_{2}$ positive between $\gamma_{2}=0.138$ and $\gamma_{2}=0.245$.
\end{itemize}

\newpage

\begin{footnotesize}
\begin{longtable}[!htb]{@{}llllllllll@{}}
    \hline
    $L/L_H$ & $L$(Gpc) & $R$ & $\langle\rho\rangle$ & median & $\delta_{1}$ & $\Sigma$ & $\delta_{1}/\Sigma$ & $\gamma_{1}$ & $\gamma_{2}$\\
    \hline
    $0.5$ & $2.2227$ & $12.57$ & $49.275$ & $49.376$ & $0.101$ & $2.723$ & $0.037$ & $-0.202$ & $-0.115$\\
    \hline
    $1.0$ & $4.4453$ & $6.29$ & $42.755$ & $42.907$ & $0.152$ & $2.45$ & $0.062$ & $-0.339$ & $0.085$\\
    \hline
    $1.5$ & $6.6680$ & $4.19$ & $41.215$ & $41.347$ & $0.132$ & $2.106$ & $0.063$ & $-0.372$ & $0.175$\\
    \hline
    $2.0$ & $8.8906$ & $3.15$ & $39.771$ & $39.858$ & $0.087$ & $1.815$ & $0.048$ & $-0.273$ & $0.085$\\
    \hline
    $3.0$ & $13.3359$ & $2.10$ & $36.173$ & $36.285$ & $0.112$ & $1.879$ & $0.060$ & $-0.356$ & $0.201$\\
    \hline
    \scriptsize{\textrm{NILC}} & $$ & $$ & \scriptsize{35.434} & $$ & $$ & $$ & $$ & $$ & $$\\
    \hline
    \scriptsize{\textrm{SEVEM}} & $$ & $$ & \scriptsize{36.290} & $$ & $$ & $$ & $$ & $$ & $$\\
    \hline
    \scriptsize{\textrm{SMICA}} & $$ & $$ & \scriptsize{35.591} & $$ & $$ & $$ & $$ & $$ & $$\\
    \hline
    \scriptsize{\textrm{C-R}} & $$ & $$ & \scriptsize{35.635} & $$ & $$ & $$ & $$ & $$ & $$\\
    \hline
    \textrm{NSSC} & $$ & $$ & $35.738$ & $35.613$ & $-0.125$ & $0.327$ & $-0.380$ & $0.971$ & $-0.781$\\
    \hline
    $\infty$ & $\infty$ & $0$ & $34.067$ & $34.143$ & $0.076$ & $1.722$ & $0.044$ & $-0.248$ & $0.035$\\
    \hline
    \caption{Table of $\rho$ (no mask), $\langle\rho\rangle$ ($\rho$ for the four Planck maps), median, $\delta_{1}$, standard deviation $\Sigma$, $\delta_{1}/\Sigma$, skewness $\gamma_{1}$ and excess kurtosis $\gamma_{2}$ 
for each of the $3-$torus side lengths and the infinite $\Lambda$CDM. NSSC stands for the ensemble of the four Planck maps NILC, SEVEM, SMICA and Commander-Ruler. 
The $3-$torus comoving side length $L$ is given in units of the Hubble length $L_H$, and $R=2r_{\hbox{\scriptsize SLS}}/L$ is twice the ratio comoving CMB 
angular diameter distance to the comoving side length of the fundamental cell, with a distance to the CMB of $r_{\hbox{\scriptsize SLS}}=14.0028$ Gpc corresponding to $3.15L_H$.}
\label{table1}
\end{longtable}
\end{footnotesize}

\begin{footnotesize}
\begin{longtable}[!htb]{@{}llllllllll@{}}
    \hline
    $L/L_H$ & $L$(Gpc) & $R$ & $\langle\rho\rangle$ & median & $\delta_{1}$ & $\Sigma$ & $\delta_{1}/\Sigma$ & $\gamma_{1}$ & $\gamma_{2}$\\
    \hline
    $0.5$ & $2.2227$ & $12.57$ & $49.542$ & $49.634$ & $0.092$ & $2.645$ & $0.035$ & $-0.184$ & $-0.109$\\
    \hline
    $1.0$ & $4.4453$ & $6.29$ & $ 43.031$ & $43.161$ & $0.130$ & $2.352$ & $0.055$ & $-0.304$ & $0.054$\\
    \hline
    $1.5$ & $6.6680$ & $4.19$ & $41.351$ & $41.471$ & $0.120$ & $2.064$ & $0.058$ & $-0.339$ & $0.138$\\
    \hline
    $2.0$ & $8.8906$ & $3.15$ & $39.622$ & $39.720$ & $0.098$ & $1.928$ & $0.051$ & $-0.274$ & $0.069$\\
    \hline
    $3.0$ & $13.3359$ & $2.10$ & $36.288$ & $36.400$ & $0.112$ & $1.888$ & $0.059$ & $-0.360$ & $0.245$\\
    \hline
    \scriptsize{\textrm{NILC}} & $$ & $$ & \scriptsize{36.639} & $$ & $$ & $$ & $$ & $$ & $$\\
    \hline
    \scriptsize{\textrm{SEVEM}} & $$ & $$ & \scriptsize{36.662} & $$ & $$ & $$ & $$ & $$ & $$\\
    \hline
    \scriptsize{\textrm{SMICA}} & $$ & $$ & \scriptsize{36.688} & $$ & $$ & $$ & $$ & $$ & $$\\
    \hline
    \scriptsize{\textrm{C-R}} & $$ & $$ & \scriptsize{36.612} & $$ & $$ & $$ & $$ & $$ & $$\\
    \hline
    \textrm{NSSC} & $$ & $$ & $36.650$ & $36.650$ & $6.9~10^{-5}$ & $2.8~10^{-2}$ & $0.002$ & $-6.2~10^{-3}$ & $-1.304$\\
    \hline
    $\infty$ & $\infty$ & $0$ & $34.132$ & $34.206$ & $0.074$ & $1.809$ & $0.041$ & $-0.244$ & $0.050$\\
    \hline
    \caption{Same as table \ref{table1} but with U73 mask.}
\label{table2}
\end{longtable}
\end{footnotesize}

\begin{footnotesize}
\begin{longtable}[!htb]{@{}llllllll@{}}
    \hline
    \multicolumn{2}{|c|}{$\boldsymbol{no~mask}$} & \multicolumn{2}{|c|}{$L/L_{H}=3$} & \multicolumn{2}{|c|}{NSSC} & \multicolumn{2}{|c|}{$\Lambda$CDM}\\  
    \hline
    \multicolumn{2}{|c|}{$\langle\rho\rangle$} & \multicolumn{2}{|c|}{$36.173$} & \multicolumn{2}{|c|}{$35.738$} & \multicolumn{2}{|c|}{$34.067$}\\
    \hline    
    \multicolumn{2}{|c|}{$\delta s$} & \multicolumn{3}{|c|}{~~~$-0.232\Sigma_{L3}$~~~} & \multicolumn{3}{|c|}{$+0.970\Sigma_{\Lambda}$}\\ 
    \hline
    \multicolumn{2}{|c|}{median} & \multicolumn{2}{|c|}{$36.285$} & \multicolumn{2}{|c|}{$35.613$} & \multicolumn{2}{|c|}{$34.143$}\\
    \hline
    \multicolumn{2}{|c|}{$\delta s$} & \multicolumn{3}{|c|}{~~~$-0.358\Sigma_{L3}$~~~} & \multicolumn{3}{|c|}{$+0.854\Sigma_{\Lambda}$}\\
    \hline
    \multicolumn{2}{|c|}{$\boldsymbol{U73~mask}$} & \multicolumn{2}{|c|}{$L/L_{H}=3$} & \multicolumn{2}{|c|}{NSSC} & \multicolumn{2}{|c|}{$\Lambda$CDM}\\
    \hline
    \multicolumn{2}{|c|}{$\langle\rho\rangle$} & \multicolumn{2}{|c|}{$36.288$} & \multicolumn{2}{|c|}{$36.650$} & \multicolumn{2}{|c|}{$34.132$}\\
    \hline
    \multicolumn{2}{|c|}{$\delta s$} & \multicolumn{3}{|c|}{~~~$+0.192\Sigma_{L3'}$~~~} & \multicolumn{3}{|c|}{$+1.392\Sigma_{\Lambda'}$}\\
    \hline
    \multicolumn{2}{|c|}{median} & \multicolumn{2}{|c|}{$36.400$} & \multicolumn{2}{|c|}{$36.650$} & \multicolumn{2}{|c|}{$34.206$}\\
    \hline
    \multicolumn{2}{|c|}{$\delta s$} & \multicolumn{3}{|c|}{~~~$+0.132\Sigma_{L3'}$~~~} & \multicolumn{3}{|c|}{$+1.351\Sigma_{\Lambda'}$}\\
    \hline
    \caption{Table of the statistical deviations $\delta s$ (see equations \eref{equ:deltasrhomedian} and \eref{equ:deltasrhomedianmask}), 
      comparing $\langle\rho\rangle$ and median of the Planck NSSC maps with the $3-$torus at $L/L_{H}=3$, also denoted $L3$ ($L3'$ with mask),
      and with the $\Lambda$CDM model, also denoted $\Lambda$ ($\Lambda'$ with mask).}
\label{table3}
\end{longtable}
\end{footnotesize}

In tables \ref{table1} and \ref{table2} one observes that the largest value for $\delta_{1}/\Sigma$ is in the no mask case $0.063$, and in the U73 mask case $0.059$,
which clearly indicates that the non-Gaussianities of $P(\rho)$ are small.\footnote{In table \ref{table2}, the very tiny values of $\Sigma$, $\delta_{1}/\Sigma$ and $\gamma_{1}$ obtained for the NSSC maps using the U73 mask are due to the fact that the observed maps constitute only one realization for a single observer position, evaluated with different pipelines of analysis. If the observations and the different pipelines would be perfect, one would obtain a zero value. So, these tiny values are a measure of the consistency of the four pipelines used by Planck in the case of the U73 mask and should not be compared with the results obtained over the ensemble of $100\, 000$ realizations (different universe models or different observer positions separated by cosmological scales) for the ${\mathcal T}^{3}$ models and the $\Lambda$CDM model.
For the same reason, the corresponding NSSC values in table \ref{table1} should not be compared with the ensemble-derived values.} 

Despite the overlap between the adjacent PDFs of each different $3-$torus, one notices that, to a given $\rho$-range, one can associate a 
given $3-$torus side length following a hierarchical ordering, i.e. the smaller the $3-$torus, the larger the $\rho$-value. In addition, the PDF of $\rho$ for the infinite 
$\Lambda$CDM model is located beyond the PDF of the largest chosen $3-$torus at $L=3.0L_H$. This trend confirms the hierarchical dependence between the size of the 
fundamental cell of the universe model and the value of the normalized standard deviation $\rho$ of the temperature gradient. Figure \ref{fig4} shows, in contrast to 
figure \ref{fig3}, the distributions obtained from the CMB maps with the application of the U73 mask, i.e. the pixels behind the U73 mask are ignored. It reveals a 
similar hierarchical ordering with the mean and median $\rho$-values somewhat shifted to higher $\rho$-values for a given torus ensemble, see also table \ref{table2}.

The two figures \ref{fig3} and \ref{fig4} also display the value of $\rho$ for each of the four foreground-corrected Planck 2015 maps, NILC, SEVEM, SMICA and Commander-Ruler.
In addition, the arithmetic average $\langle\rho\rangle$ for these four Planck maps (NSSC) is shown (see tables \ref{table1} and \ref{table2}). 
Their individual $\rho$-values are indicated by the four vertical lines in the two plots. These $\rho$-values can be clearly distinguished in figure \ref{fig3},
where the foreground-contaminated pixels are present. These $\rho$-values, however, nearly converge to the arithmetic average $\langle\rho\rangle$, when the U73 mask pixels are rejected, as can be appreciated in figure 
\ref{fig4}. The arithmetic average $\langle\rho\rangle = 35.738$ of the four Planck maps is rather close to the arithmetic average $\langle\rho\rangle \sim 36.173$ of the $3-$torus ensemble $L=3.0L_H$ at $-0.232 \Sigma_{L3}$ 
(see equations \eref{equ:deltasrhomedian} and \eref{equ:deltasrhomedianmask} for definition of the statistical deviations) when no mask is used, see table \ref{table1},
and, with the U73 union mask, the arithmetic average $\langle\rho\rangle = 36.650$ of the four Planck maps is $+0.192\Sigma_{L3'}$ above the arithmetic average at $36.288$ of the $3-$torus sample $L=3.0L_H$, see table \ref{table2}.

Without mask (see table \ref{table1}), the median value $35.613$ of the four Planck maps is slightly below the median at $36.285$ of the $3-$torus ensemble, 
i.e. at $-0.358 \Sigma_{L3}$. With the U73 mask (see table \ref{table2}), the median of the NSSC maps at $36.650$ is a little above, i.e. at $+0.132 \Sigma_{L3'}$ of the median $36.400$ of the $3-$torus sample $L=3.0L_H$.
These results of the statistical deviation $\delta s$ of $\langle\rho\rangle$ and median for the four NSSC Planck maps compared with the $3-$torus at $L=3.0L_H$ are shown in the synoptic 
table \ref{table3}. This table applies the same method to compare the NSSC maps with the $\Lambda$CDM maps, and we discuss these further results at the end of section \ref{sec:conclusion}.

\begin{figure}[!htb]
\includegraphics[height=0.53\textheight,width=0.53\textwidth, angle=270]{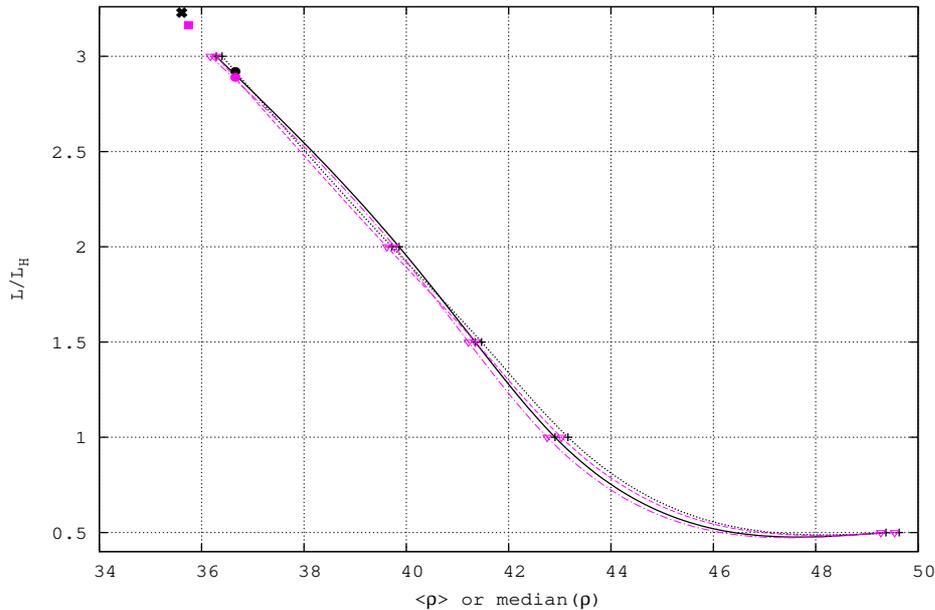}
\caption{The side length $L/L_H$ of the $3-$torus as a function of the median of $\rho$ in black and of $\langle\rho\rangle$ in magenta for the online version (in black for the print version), 
for $L/L_H= [0.5, 1.0, 1.5, 2.0, 3.0]$.
The median case without mask is in black solid line, and in black small-dotted line with U73 mask; the $\langle\rho\rangle$ case without mask is in dotted-dashed line, and in small-dashed line with U73 mask. 
For the median case, the thick black cross in the upper left at $L/L_H\approx 3.229$, respectively the thick black dot at $L/L_H\approx 2.920$, point at the 
side length of the ${\mathcal T}^3$ estimated from equation \eref{equ:Lmednomsk}, respectively equation \eref{equ:LmedU73}, using as argument the $\langle\rho\rangle$ of the NSSC Planck maps without 
mask, respectively with U73 mask.
Similarly, for the $\langle\rho\rangle$ case of the Planck maps, the equation \eref{equ:Lavernomsk} without mask points at a side length of $L/L_H\approx 3.164$ (the solid square in the upper left), 
respectively the equation \eref{equ:LaverU73} with U73 mask pointing at a side length of $L/L_H\approx 2.889$ (the thick inferior dot).  
}  
\label{fig5}
\end{figure}

The statistical deviation $\delta s$ of the NSSC ensemble (denoted NSSC' with mask) in comparison with the $3-$torus at $L/L_{H}=3$ (denoted $L3$ or $L3'$ with mask) 
or the $\Lambda$CDM model ensembles (denoted $\Lambda$ or $\Lambda'$ with mask) is defined the following way without mask:
\begin{equation}
  \label{equ:deltasrhomedian}
  $$
        {\delta s}:=\left\{
        \begin{array}{ll}
          \frac{\langle\rho\rangle_{\textrm{\scriptsize NSSC}} - \langle\rho\rangle_{L3}}{\Sigma_{L3}} & \mbox{\scriptsize{, for $\langle\rho\rangle$ and the $3-$torus at $L/L_{H}=3$}}\\
          \frac{\langle\rho\rangle_{\textrm{\scriptsize NSSC}} - \langle\rho\rangle_{\Lambda}}{\Sigma_{\Lambda}} & \mbox{\scriptsize{, for $\langle\rho\rangle$ and $\Lambda$CDM}}\\
          \frac{\textrm{\scriptsize median}_{\textrm{\scriptsize NSSC}} - \textrm{\scriptsize median}_{L3}}{\Sigma_{L3}} & \mbox{\scriptsize{, for the median and the $3-$torus at $L/L_{H}=3$}}\\
          \frac{\textrm{\scriptsize median}_{\textrm{\scriptsize NSSC}} - \textrm{\scriptsize median}_{\Lambda}}{\Sigma_{\Lambda}} & \mbox{\scriptsize{, for the median and $\Lambda$CDM}}\;\;,
        \end{array}
        \right.
        $$
\end{equation}
and with U73 mask:
\begin{equation}
  \label{equ:deltasrhomedianmask}
  $$
        {\delta s}:=\left\{
        \begin{array}{ll}
          \frac{\langle\rho\rangle_{\textrm{\scriptsize NSSC}'} - \langle\rho\rangle_{L3'}}{\Sigma_{L3'}} & \mbox{\scriptsize{, for $\langle\rho\rangle$ and the $3-$torus at $L/L_{H}=3$}}\\
          \frac{\langle\rho\rangle_{\textrm{\scriptsize NSSC}'} - \langle\rho\rangle_{\Lambda'}}{\Sigma_{\Lambda'}} & \mbox{\scriptsize{, for $\langle\rho\rangle$ and $\Lambda$CDM}}\\
          \frac{\textrm{\scriptsize median}_{\textrm{\scriptsize NSSC}'} - \textrm{\scriptsize median}_{L3'}}{\Sigma_{L3'}} & \mbox{\scriptsize{, for the median and the $3-$torus at $L/L_{H}=3$}}\\
          \frac{\textrm{\scriptsize median}_{\textrm{\scriptsize NSSC}'} - \textrm{\scriptsize median}_{\Lambda'}}{\Sigma_{\Lambda'}} & \mbox{\scriptsize{, for the median and $\Lambda$CDM}}\;\;.
        \end{array}
        \right.
        $$
\end{equation}
The $\rho$-statistics is thus favouring a $3-$torus size slightly larger than $3 L_H$ in the case without mask and 
is consistent with a $3-$torus of side length $3L_{H}\approx 13.336\,$Gpc in the case with U73 mask. 
The analysis of $\rho$ median and $\langle\rho\rangle$ with respect to the $3-$torus side length $L$ clearly shows 
(see the figures \ref{fig3} and \ref{fig4}) that the derivatives are negative, ${\mathrm d}(\textrm{median})/{\mathrm d}L   < 0$ and  ${\mathrm d}\langle\rho\rangle / {\mathrm d}L < 0$,
as it is quantified by the linear equations \eref{equ:Lmednomsk}, \eref{equ:Lavernomsk}, \eref{equ:LmedU73} and \eref{equ:LaverU73} obtained by linear least square fitting (thereafter LSF).
Figure \ref{fig5} shows the relation between the side length $L$ of the cubic $3-$torus and the median or the arithmetic mean of $\rho$ obtained from the samples consisting of $100\,000$ maps.

Except below $L=1.0L_H$, the curves of $L=f($median$)$ and $L=f(\langle\rho\rangle)$ look close to linear between $L=1.0L_H$ and the three larger side lengths up to $L=3.0L_H$.
In the case without a mask,
the linear least square fitting for the median case in the interval $36.285 \le \textrm{median}_{\hbox{\scriptsize nomask}}\le 42.907$ yields
\begin{equation}
\label{equ:Lmednomsk}
{ \frac{ L_{\hbox{\scriptsize nomask}}(\textrm{median}) }{L_H} } \approx-0.302~\textrm{median}_{\hbox{\scriptsize nomask}} + 13.981\;, \end{equation}
and for the $\langle\rho\rangle$ case in the interval $36.173 \le \langle\rho\rangle_{\hbox{\scriptsize nomask}}\le 42.755$,
the LSF gives 
\begin{equation}
\label{equ:Lavernomsk}
{ \frac{ L_{\hbox{\scriptsize nomask}}(\langle\rho\rangle) }{L_H} } \approx-0.304~\langle\rho\rangle_{\hbox{\scriptsize nomask}} + 14.021\;. \end{equation}
With applying the U73 mask, 
the LSF for the median case in the interval $36.400 \le$median$_{\scriptsize{\textrm{U73}}}\le 43.161$ yields
\begin{equation}
\label{equ:LmedU73}
{ \frac{ L_{\hbox{\scriptsize U73}}(\textrm{median}) }{L_H} } \approx-0.295~\textrm{median}_{\scriptsize{\textrm{U73}}} + 13.751\;\;, \end{equation}
and for the $\langle\rho\rangle$ case in the interval $36.288 \le \langle \rho\rangle_{\hbox{\scriptsize U73}}\le 43.031$, the LSF gives 
\begin{equation}
\label{equ:LaverU73}
{ \frac{ L_{\hbox{\scriptsize U73}}(\langle\rho\rangle) }{L_H} } \approx-0.296~\langle\rho\rangle_{\hbox{\scriptsize U73}} + 13.750\;. \end{equation}
One may visually observe in figure \ref{fig5} the better agreement with the linear behaviour of the curves with U73 mask 
(small dotted line for the median-case or small dashed line for the $\langle\rho\rangle$-case) 
in comparison to the slightly twisted curve (solid line or dotted dash line for the $\langle\rho\rangle$-case) obtained without mask pixel suppression. 
The $\chi^{2}$ comparing the data points to the LSF's being with U73 mask $1.72 ~ 10^{-6}$ for the median and $7.5 ~ 10^{-7}$ for $\langle\rho\rangle$, while 
the $\chi^{2}$ without mask is $2.929 ~ 10^{-5}$ for the median and $4.040 ~ 10^{-5}$ for $\langle\rho\rangle$. These $\chi^{2}$-values are $\sim$17 (for the median) and $\sim 54$ (for $\langle\rho\rangle$) times larger without mask than with U73 mask. 
Thus, given the median and average $\rho$ values 
of the four Planck NSSC maps, these LSF's of the data points yield, with the hypothesis of a flat $3-$toroidal topology of our Universe, a side length between $2.89$ and $3.23$
($3.16\le L/L_H$ (no mask) $\le3.23$ and $2.89\le L/L_H$ (U73 mask) $\le 2.92$). 

According to the works \cite{planck4,bernui1,fabre1}, $3-$torus side lengths that
are barely bigger than the CMB diameter
($L_{\hbox{\scriptsize limit}}=2.2 r_{\hbox{\scriptsize SLS}}$,
which translates to $6.93L_H$, corresponding to a threshold ratio $R=0.91$),
do not allow for a clear detection of a multiply connected topology in the sense of the Kullback--Leibler divergence. 
A reasonable spatial section size that results in no difference with the infinite Universe was proposed in \cite{bernui1} to be $L_{\infty}=4r_{\hbox{\scriptsize SLS}}$ = $12.6L_H$ giving $R = 0.5$. For this paper we did not calculate $\rho$-values for $L$ bigger than three Hubble radii to analyze the asymptotic behaviour of $L=f(\rho)$ presented in figure \ref{fig5}.

\section{Comparison of two tori: $\mathbf{L=0.5\,L_H}$ and $\mathbf{L=3.0\,L_H}$}
\label{sec:comparison}

\begin{figure}[!htb]
\begin{minipage}[b]{1.0\linewidth}
\includegraphics[height=0.35\textheight,width=1.0\textwidth, angle=0]{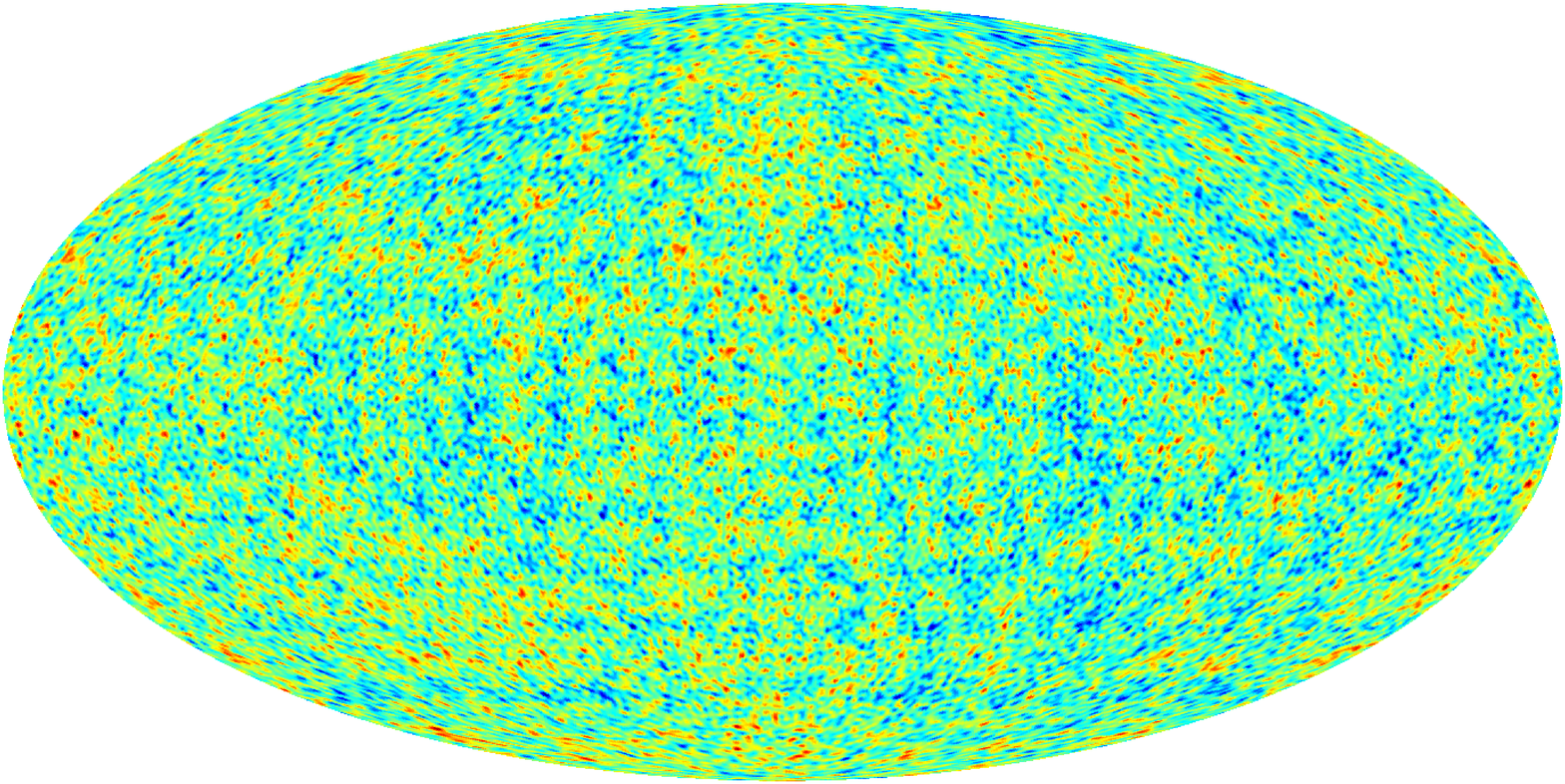} 
\caption{This figure shows a simulated CMB sky map (having the monopole and dipole subtracted) for a small cubic $3-$torus fundamental cell of $L=0.5L_H$.
The resolution parameters are $N_{\hbox{\scriptsize side}} =256$, $l_{\hbox{\scriptsize range}}=[2,256]$
and $\vartheta_{G} =(1/3)^{\circ}$.\\}
\label{fig6}
\end{minipage}
\vspace{20pt}
\begin{minipage}[b]{1.0\linewidth}
\includegraphics[height=0.35\textheight,width=1.0\textwidth, angle=0]{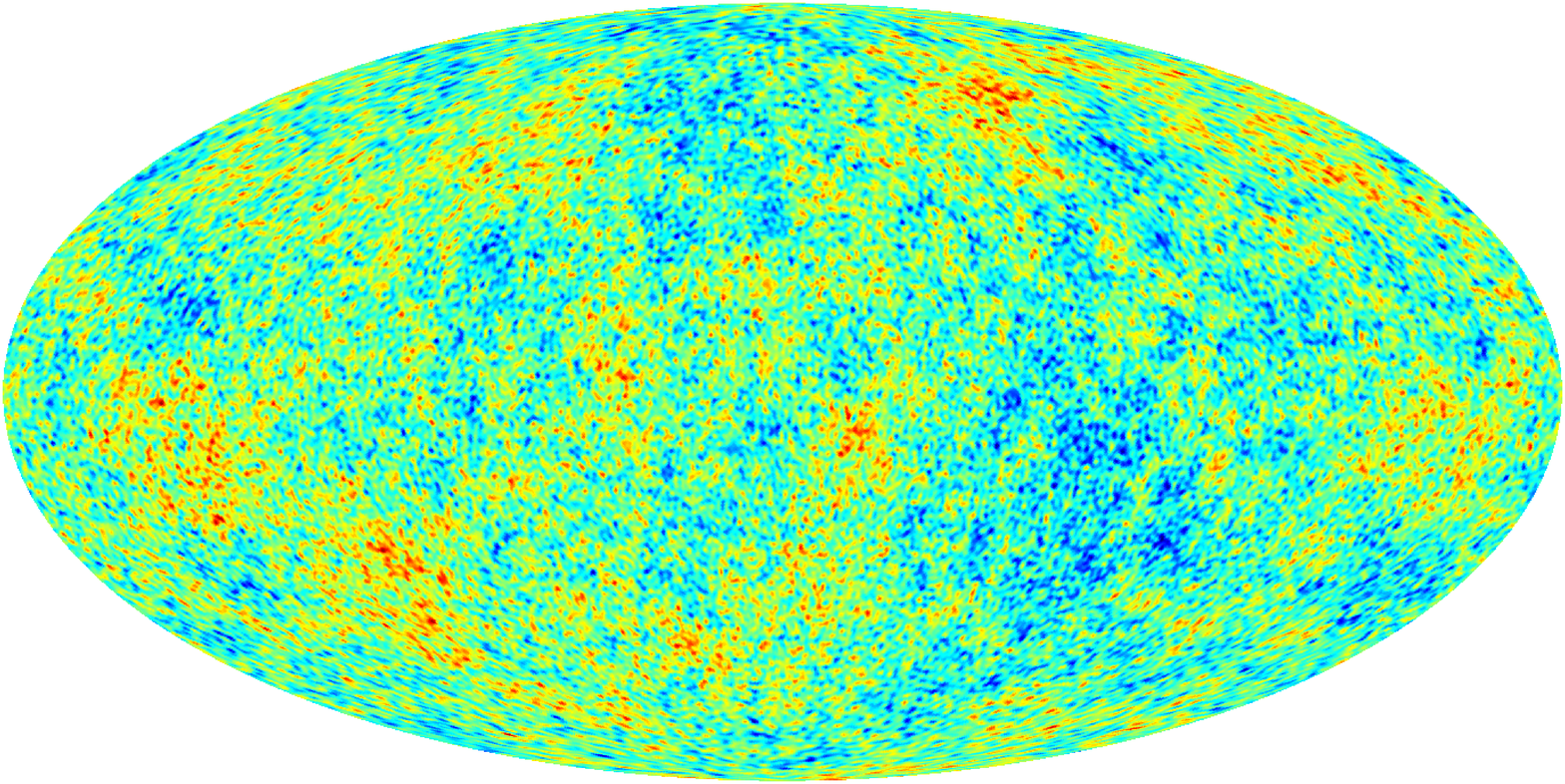} 
\caption{Figure showing a simulated CMB sky map (having the monopole and dipole subtracted) for a cubic $3-$torus fundamental cell of $L=3.0L_H$,
which is six times larger than the one used for figure \ref{fig6}. 
The resolution parameters are those of the figure \ref{fig6}.}
\label{fig7}
\end{minipage}
\end{figure}

A CMB map for a $3-$torus topology at $L=0.5L_H$ is shown in figure \ref{fig6} and reveals that the small--scale structures are dominant, i.e. the anisotropy gradients at the smallest scales are strong almost everywhere, while no obvious structure at large scales appears.
This contrasts to the CMB map for a six times larger $3-$torus at $L=3.0L_H$ (figure \ref{fig7}), where the small--scale structures are superposed 
by large--scale structures, i.e. larger areas with similar temperatures are patching the CMB map.
This is caused by the decreasing suppression of large--scale fluctuations with increasing size of the fundamental cell, which is also revealed by 
the multipole spectrum $C_l$ or the 2--pcf $C(\vartheta)$. The small smoothing scale of $\vartheta_{G} =(1/3)^{\circ}$, which is applied in the CMB maps shown in figures
\ref{fig6} and \ref{fig7}, does not influence those features. 

\begin{figure}[!htb]
\begin{minipage}[b]{1.0\linewidth}
\includegraphics[height=0.6\textheight,width=0.6\textwidth, angle=270]{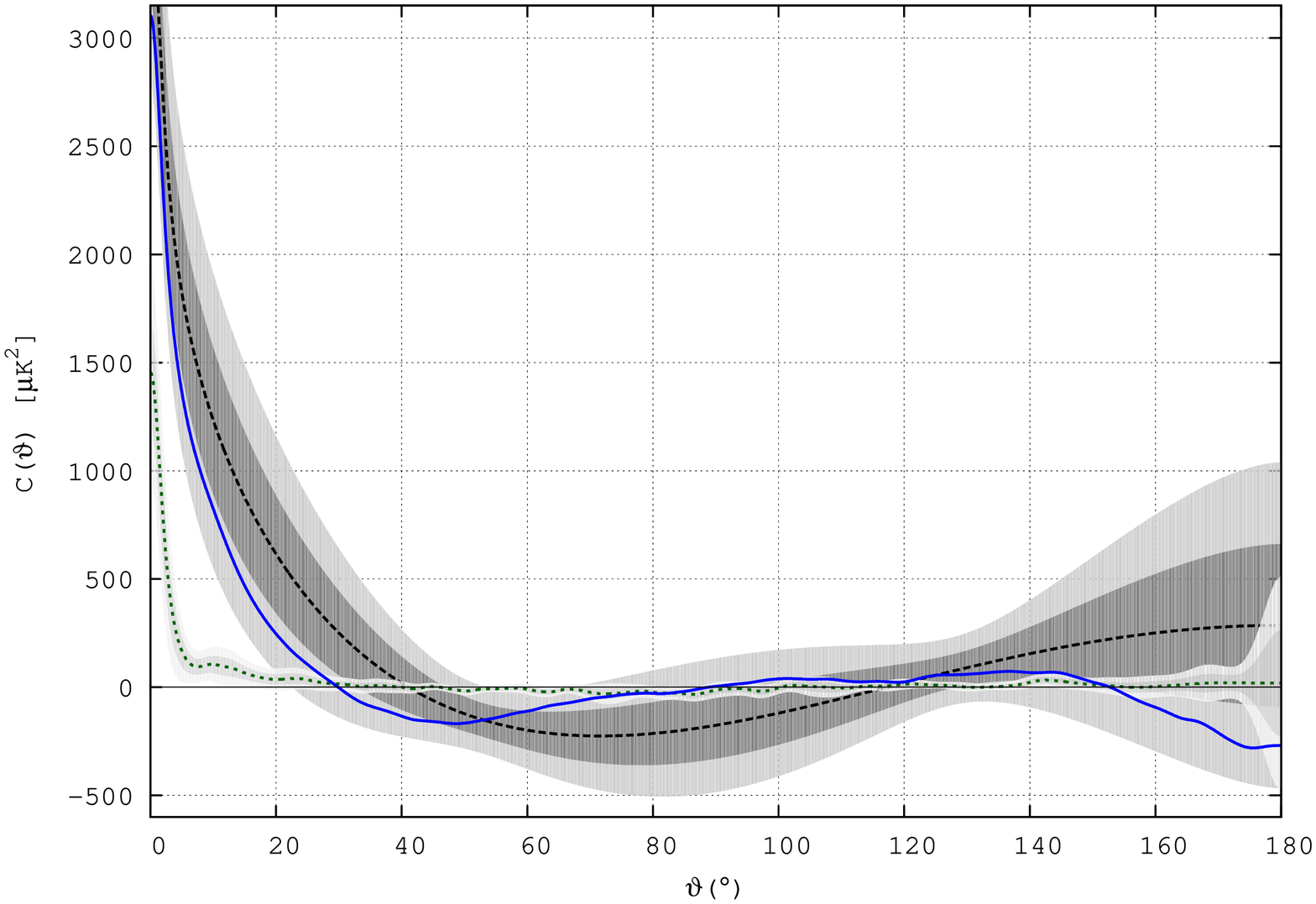}
\caption{For $100\,000$ CMB maps without mask, at $N_{side}=128$, l$_{max}=256$ and a Gaussian smoothing of $2^{\circ}$fwhm: the average 
two-point correlation functions of the $\Lambda$CDM ensemble in large-dashed line (black in the online version), 
of the torus at $L=0.5L_H$ in green small-dashed line; $\pm 1\sigma$ in dark shaded area and $\pm 2\sigma$ in light shaded area are shown versus the average 2--pcf of the four Planck NSSC maps in solid line (blue).\\}
\label{fig8}
\end{minipage}
\begin{minipage}[b]{1.0\linewidth}
\includegraphics[height=0.6\textheight,width=0.6\textwidth, angle=270]{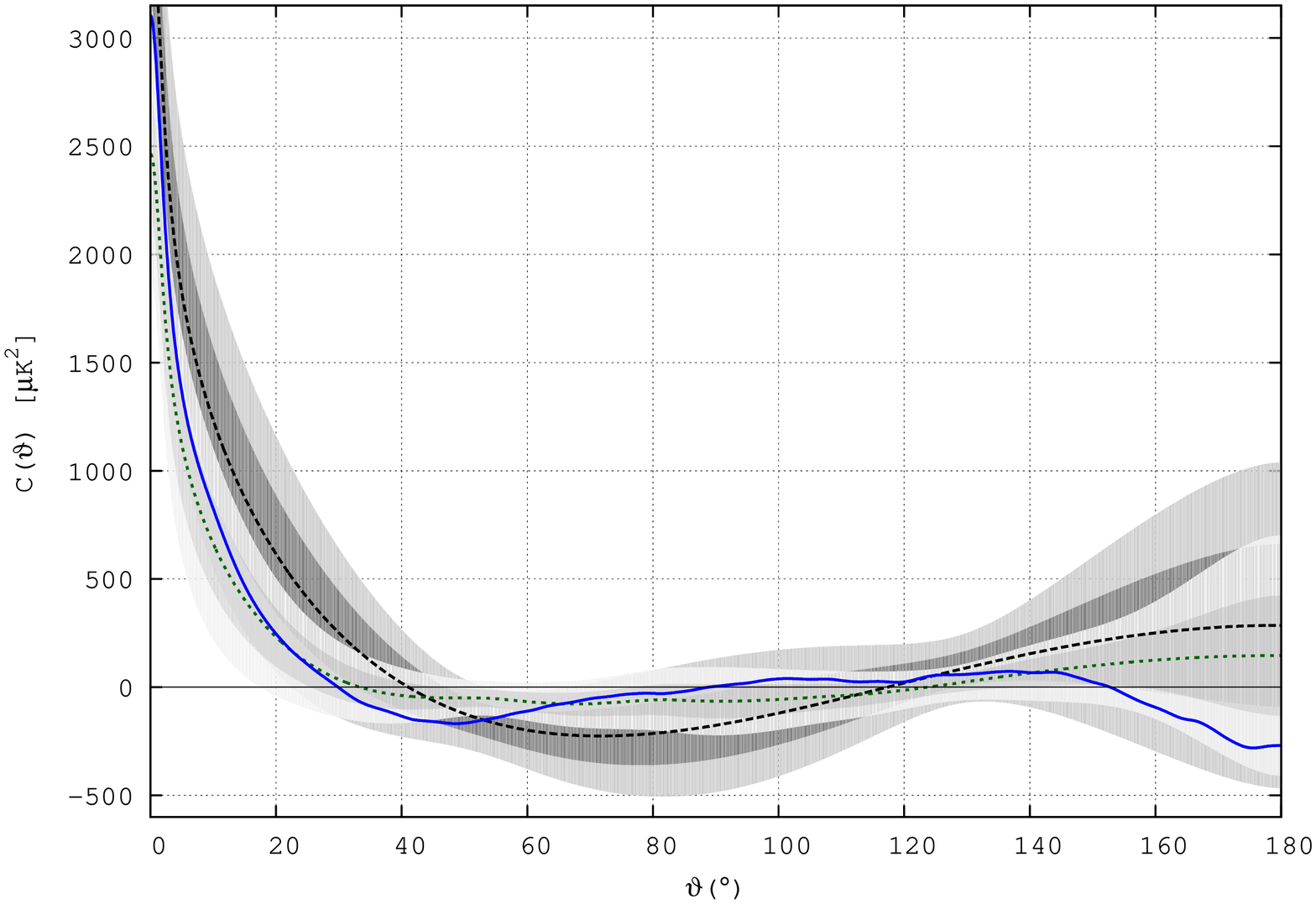}
\caption{Same as figure \ref{fig8}, but for the torus at $L=3.0L_H$.}
\label{fig9}
\end{minipage}
\end{figure}

In both cases, a scale typical for the underlying $3-$torus size visually betrays the topology (see the 2--pcf signature of each of these side lengths in figures \ref{fig8} and \ref{fig9}).
This visual illustration is in accordance with the conclusions in section~\ref{sec:dependence} that the normalized local
CMB gradient $\rho$ characterizes and quantifies the $3-$torus side length.\\

The CMB maps of different $3-$torus sizes and of the infinite $\Lambda$CDM model have to be normalized in order to get the first 
acoustic peak of the power spectrum at the same level as in the Planck observation map. To this aim, the transfer function is computed for each averaged torus model, and the $1^{st}$ 
acoustic peak of the corresponding $C_{l}$ spectrum is fitted to the $1^{st}$ peak of the Planck spectrum. 

Figure \ref{fig8} (respectively figure \ref{fig9}) display, for the case without mask, the average 2--pcf (over $100\,000$ simulation maps) of the 
torus at $L=0.5L_H$ (respectively at $L=3.0L_H$), compared with the average 2--pcf of the ensemble of $100\,000$ $\Lambda$CDM simulation maps and to the average 2--pcf of the four NSSC Planck maps.

An examination of the 2--pcfs of the cubic torus with $L=0.5$ (shown in figure~\ref{fig8}), and $3.0L_H$ (shown in figure \ref{fig9}) reveals the following:
\begin{itemize}
\item[--] the torus with $L=$0.5$L_H$ has no correlation for the pairs of pixels separated by more than $30^{\circ}$, on average;
\item[--] between $10^{\circ}$ and $30^{\circ}$, and between $60^{\circ}$ and $145^{\circ}$, the average 2--pcf for $L=3.0L_H$ fits well the average 2--pcf of the Planck NSSC, better than the average 2--pcf of the 
$\Lambda$CDM model;
\item[--] for the small angles $\vartheta$ below $30^{\circ}$ also the $\pm 1\sigma$ confidence region of the $3-$torus at $L=3.0L_H$ does not overlap with the corresponding region of the $\Lambda$CDM model.
\end{itemize}

\newpage

\newpage

\section{Isotropy and homogeneity of the CMB with toroidal topology}
\label{sec:torusisotropy}

\begin{figure}[!htb]
\includegraphics[height=0.60\textheight,width=0.60\textwidth, angle=270]{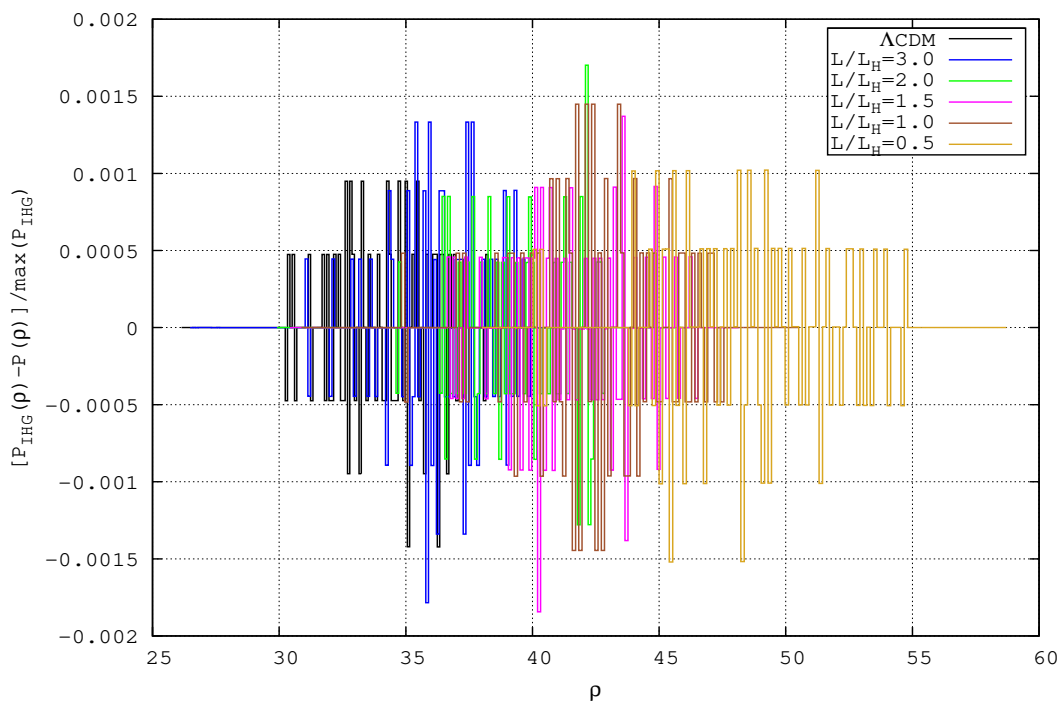}
\caption{The level of isotropy and homogeneity of the CMB in a Universe with $3-$torus topology is quantified with the discrepancy functions $\Delta P(\rho)$ of the histogram of $P(\rho)$ without mask. 
This figure is presented in solid lines for the online version with from left to right the $\Lambda$CDM in black, the $3-$torus at $3.0 L_H$ in blue, $2.0 L_H$ in green, $1.5 L_H$ in magenta, $1.0 L_H$ in brown and at $0.5 L_H$ in light brown.}
\label{fig10}
\end{figure}

We define a discrepancy function of the histogram of $P(\rho)$ shown in figure \ref{fig3} by
\begin{equation}
\label{equ:discrfctpdfrho}
{\Delta P(\rho)}:=\frac{P_{\textrm{\scriptsize IHG}}(\rho)-P(\rho)}{\textrm{\small max}(P_{\textrm{\scriptsize IHG}})}\;\;,
\end{equation}
where the histogram $P_{\textrm{\scriptsize IHG}}(\rho)$ is determined from the equations \eref{equ:sigma1sq5} and \eref{equ:sigma0sq5}, while the histogram $P(\rho)$ is determined using equations  
\eref{equ:sigma1sq3} and \eref{equ:sigma0sq}. This quantifies the drift of the $3-$torus CMB maps from the hypothesis of isotropy and homogeneity. We present 
in figure \ref{fig10} the shape of the function \eref{equ:discrfctpdfrho} for the map ensembles of the $\Lambda$CDM and the $3-$torus. 

The discretization of figure \ref{fig10} is due to the very close values taken by the two histograms in each bin so that the discrepancy function progresses by leaps, because 
the histograms with IHG or without IHG differ only by zero or by a few multiples of unity before normalization. 
Despite the large number of $100\,000$ maps used for this $\rho$-statistics, figure \ref{fig10} does neither present a smooth behaviour nor shape similarities from one $3-$torus to another.
Finally, this test proves the extremely high level of isotropy and homogeneity (in the sense of the formulas \eref{equ:sigma1sq5} and \eref{equ:sigma0sq5}) of all the
ensembles of maps. 
This test over $100\,000$ maps allows to draw a firm conclusion, confirming that the $\Lambda$CDM CMB map ensemble is closer to the perfect IH
(this is not a test of IHG but only of IH). 
The violation of the isotropy in the sense of $\rho$ is nearly as small as for the five CMB map ensembles of the $3-$torus under scrutiny given that $\vert \Delta P(\rho) \vert < 0.18\%$ 
for all the map ensembles. 
Thus, the relative global anisotropy of the $3-$torus models barely appears here and we will come later to methods able to detect it.    
Very likely, a similar analysis applied to the same large sample sizes but with a higher spatial resolution would lead to the same weak anisotropy and inhomogeneity.

\section{Discussion}
\label{sec:discussion}

The $3-$torus simulations of the CMB temperature anisotropies were computed by 
implementing the following effects of the Boltzmann physics and the influence of the discrete spectrum of vibrational modes dictated by the topology:
the roster of physical ingredients of the $3-$torus simulations includes the ordinary and integrated Sachs-Wolfe effects,
the Doppler effect, Silk damping, reionization, photon polarization and neutrinos. The computation of the CMB anisotropies (CMB power spectrum) is carried out along the lines presented in \cite{aurich17}.
We use as in \cite{planck5} the definition of low-$l$ values $l\in[2,29]$ (see e.g. their figures 2 and 3 on page 6 of \cite{planck5}) and high-$l$ values for $l\ge30$. At high-$l$ values, the angular power spectrum 
$\delta T_{l}^{2}:=l(l+1)C_{l}/2\pi$ gets smoother and smoother and approaches for instance, near the first acoustic peak at $l=221$ and for all the different $3-$torus side lengths, the 
$\Lambda$CDM result (shown in \cite{planck5}, figure 57).

For the CMB simulations in the $\Lambda$CDM model, in addition to the effects enumerated above, lensing is present too. However, the impact of lensing would be sensitive for maps 
with $l_{max} \ge 400$ (see \cite{durrer1}) but all of our maps are limited to $l_{max}=256$, 
and are furthermore smoothed to a resolution of $2^{\circ}$ f.w.h.m. Because of this smoothing we have almost no power above $l=100$ ... $150$. 
Thus, the comparison between the $3-$torus and the $\Lambda$CDM CMB maps remains unaffected by the effect of weak lensing in the $\Lambda$CDM simulation maps. 
Fully accounting for all these effects in a universe model with multiply or even simply connected 
topology for an \textit{analytic} prediction of CMB observables such as $\rho$, or for a statistics such as the 2--pcf, is for the moment out of reach.
The $\sigma_{n}$'s defined for $n=0,1,2, ...$ by\cite{aurich14},
\begin{equation}
\label{equ:sigmas}
{\sigma_n^2}:=\sum_{l=2}^{\infty} \frac{2l+1}{4\pi} C_l |F_l|^2 \frac{(l+n)!}{(l-n)!}\;\;,
\end{equation}
are decreasing functions of $\vartheta_{G}$, the scale of Gaussian smoothing (full width at half maximum), defined in equation \eref{equ:gausskernel}.
However, the decrease of \eref{equ:sigmas} does not imply that the normalized variance of the gradient field, $\rho=\sqrt{\sigma_1^2/\sigma_0^2}$, of a CMB map 
is also everywhere a decreasing function of the smoothing angle $\vartheta_G$. 

In a flat Universe having three infinite spatial directions such as the $\Lambda$CDM model the spectrum is continuous. 
The average 2--pcf of the CMB map sample in the $\Lambda$CDM model (large-dashed line 
e.g.~in figure \ref{fig9}) shows correlations at all angular scales. 

\section{Conclusion and Outlook}
\label{sec:conclusion}
Our investigation shows that $\rho$ is a powerful signature probe that is sensitive to the size and the compactness of the spatial sections of the Universe. 
The $\rho$-statistics allows to hierarchically discriminate compact fundamental cells having the same $3-$torus topology but different volumes. A clear distinction between a 
multiply connected flat universe model (the cubic $3-$torus) and a simply connected flat universe model with infinite spatial sections (the $\Lambda$CDM model) is nicely verified for torus side lengths smaller than $L \sim 3L_{H}$ as shown in figures \ref{fig3} and \ref{fig4}. For tori larger than about $L=3.0L_{H}$ (see the discussion at the end of section \ref{sec:dependence}), the calculation of a more refined grid of models would be needed. 
Different observables allow to detect a given multiply connected topology in a different way. On the one hand, the 2--pcf is able to detect on the CMB map the different 
angular scales and the size of a given fundamental domain. On the other hand, $\rho$ is by definition extremely sensitive to a change of the normalized CMB gradient as a function of the domain size and of the smoothing angle and amplitude. 
The $\rho$-statistics furnishes a complementary test of the multiply connected nature of the Universe along with the 2--pcf. The present results based 
on samples of $100\,000$ CMB maps with cubic torus topology are consistent with a Dirichlet domain side length of our Universe of $\sim 3.20 L_H$, 
or $\sim 2.90 L_H$ when the Planck and torus maps are cleaned up from the contaminated mask pixels. 
The investigated $\rho$-test may be included in
future Bayesian analyses of model selection, with the expectation that a 
torus model  in the above size range with the U73 mask might be favoured over the flat infinite model. 

For the Planck maps, the $3-$torus size around three Hubble radii or below, inferred from this $\rho$-study, is therefore slightly smaller than the torus size of $3.69 L_H$ inferred from the 2--pcf investigations. 
It remains to be seen whether other statistics like the Minkowski functionals, may lead to a slightly different optimal torus size. The sources of such a difference as well as the systematically negative skewness are currently investigated in a theoretical model for the PDF $P(\rho)$ \cite{aurich21} and thoroughly probed in other projects that employ the Minkowski functionals and topological characterization using Betti numbers and homological concepts such as hierarchical persistence, e.g. \cite{pratyush1}. 
It will have to be verified that $\rho$ could more generally detect size changes in finite fundamental cells of any geometry and topology.  The vibrational modes (wave numbers and eigenfunctions of the Laplacian) along each compact spatial section and the interference (destructive or constructive) of these vibrational modes reveal the possible shapes of the underlying topological 
manifold. The 2--pcf says nothing about the non--Gaussianity of a random field. 
For some compact manifolds there are analytic premises of the CMB 2--pcf for the Sachs-Wolfe contribution, e.g. for the Poincar\'e dodecahedron\cite{aurich8}, and general spherical spaces\cite{aurich8,aurich2}. 
Thus, the 2--pcf and $\rho$ lead to identical diagnoses in two conceptually different ways.

The possibility of detecting the circle-in-the-sky (CITS) signal of a multiply connected topology has been discussed at the end of section~\ref{introduction}. It is a geometric signal, while the 2--pcf and $\rho$ are statistical observables.
In \cite{planck4}, $\Lambda$CDM temperature simulation maps (with noise and Gaussian smoothing of the Planck SMICA map) with cubic $3-$torus topology at $L/L_H=2$ present all the pairs of matched circles that are expected. Also for the associated simulation maps of E-mode polarization, the multiply connected topology is detected with the $S^-_{\rm max}(\alpha)$ statistics. However, the same detection tools applied to the Planck 2015 observation maps for different circle patterns show no evidence of multiply connected topology with a size smaller than the distance to the CMB, i.e. $L \approx 3.15L_H$. In view of recent survey results regarding the strong ISW effect, it is possible that this ISW signal 
(which is stronger than in the $\Lambda$CDM $3-$torus simulations) 
impairs the detection of the CITS signal. 

\vspace{20pt}

\noindent
{\em Acknowledgements:}
\small{This work is part of a project that has received funding from the European Research Council (ERC) under the European Union's Horizon 2020 research and innovation programme (grant agreement ERC adG No.\ 740021--ARThUs, PI: TB). FS is grateful to Sven Lustig for collaboration at an early stage of this work.
The authors wish to thank Robert J Adler, L\'eo Brunswic, Neil Cornish, Pratyush Pranav and Quentin Vigneron for valuable discussions and remarks, and the anonymous referees for their useful comments. We gratefully acknowledge support from the PSMN (P\^ole Scientifique de Mod\'elisation 
Num\'erique) of the \'Ecole Normale Sup\'erieure de Lyon for the computing resources [\href{http://www.ens-lyon.fr/PSMN/doku.php?id=en:accueil:presentation}{PSMN presentation}].
This work is also based on observations obtained with Planck [\href{http://www.esa.int/Planck}{Planck}], an ESA science mission with instruments and 
contributions directly funded by ESA Member States, NASA, and Canada. Some of the results in this paper have been derived using the HEALPix package \cite{gorski1}, available 
at [\href{https://healpix.jpl.nasa.gov/}{HEALPix}].
The CMB power spectra of the infinite $\Lambda$CDM model are calculated from the cosmological parameters using $CAMB$ software written by Lewis and Challinor
[\href{http://lambda.gsfc.nasa.gov/toolbox/tb_camb_form.cfm}{CAMB interface}] from the original Boltzmann codes by Bertschinger, Ma and 
Bode resumed by Seljak and Zaldarriaga. The $CAMB$ ReadMe (2016) is available here [\href{http://camb.info/readme.html}{CAMB readme}] and $CAMB$ Notes by 
A Lewis (2014) here [\href{http://cosmologist.info/notes/CAMB.pdf}{CAMB Notes}].\\}

\appendix

\section{The truncated Gaussian probability density function}
\label{appA}

The construction of the truncated Gaussian PDF is based on the standard unconditional normal distribution (also called in this context the ``parent distribution'' by statisticians)
defined on the whole line in terms of the mean $\mu$ and the variance $\sigma^{2}$. 
The \textit{truncated Gaussian} PDF $P_{+}(\rho)$ 
is defined as the normalized conditional PDF restricted to the half-line $[0,\infty[$ by 
\begin{equation}
\label{equ:pplus}
{P_{+}(\rho)}:=\frac{N}{\sqrt{2\pi}\sigma} \exp{\left(-\frac{(\rho-\mu)^{2}}{2\sigma^{2}}\right)} \Theta(\rho)\;\;,
\end{equation}
where $\Theta(\rho)$ is the Heaviside step function ($\Theta(\rho)=1$ for $\rho\ge0$, $\Theta(\rho)=0$ for $\rho<0$). $N$ is a normalization constant determined by the parent parameters
$\mu>0$ and $\sigma>0$ and is given by
\begin{equation}
\label{equ:N}
{N}:=\frac{2}{ 1+ \textstyle{\rm erf}(\frac{\mu}{\sqrt{2}\sigma}) }\;\;,
\end{equation}
satisfying $1<N<2$. It follows that $P_{+}(\rho)$ is unimodal of mode $\mu$ having the same shape as the standard normal distribution whose peak height $b$ at $\rho=\mu$ is,
however, larger by the factor $N$. The important new properties of $P_{+}(\rho)$ are that the mean $\langle\rho\rangle$ is no more equal to the mode $\mu$ and is also different from the median, and that the variance
$\Sigma^{2}$ is different from the parent variance $\sigma^{2}$. Actually, all higher moments are different from the 
well-known Gaussian moments, in particular the odd moments and all higher cumulants are non-zero. 
As an example we give the values for $\langle\rho\rangle$ and $\Sigma^{2}$:
\begin{equation}
\label{equ:rhoaver}
{\langle\rho\rangle}=\mu+\sigma \lambda > \mu\ ;
\end{equation}
\begin{equation}
\label{equ:Sigmasq}
{\Sigma^{2}}:= \left\langle(\rho-\langle\rho\rangle)^{2}\right\rangle=\sigma^{2}-\sigma^{2} \lambda \left(\lambda+\frac{\mu}{\sigma}\right) < \sigma^{2}\ ,
\end{equation}
with
\begin{equation}
\label{equ:lambda}
{\lambda}:=\frac{N}{\sqrt{2\pi}} \exp{\left( -\frac{\mu^{2}}{2\sigma^{2}}\right)} \;\;.
\end{equation}
Note that $\langle\rho\rangle$ and $\Sigma^{2}$ can be directly computed from the histograms. But in order to compare the histograms with the truncated (continuous) Gaussian $P_{+}(\rho)$,
one has to know the \textit{a priori} unknown parent parameters $\mu$ and $\sigma$. In principle, it is straightforward to get them. $\mu$ is directly determined by the position of the maximum of the histogram, 
and $\sigma$ is obtained from the numerical solution of the equation (see equations \eref{equ:pplus} and \eref{equ:N}),
\begin{equation}
\label{equ:sigmaparent}
{\sigma+\sigma\, \textstyle{\rm erf}\left(\frac{\mu}{\sqrt{2}\sigma}\right)}=\sqrt{\frac{2}{\pi}}~\frac{1}{b}\;\; ,
\end{equation}
once the previously obtained value for $\mu$ and the peak height $b$ have been inserted. The actual determination of $\mu$ and $\sigma$ is, however, rendered more difficult, since the
accuracy of the histograms depends on the binning and, thus, the correct position of the maximum (and of $b$) is not well-defined (see figures \ref{fig3} and \ref{fig4}).

From equations \eref{equ:N}--\eref{equ:sigmaparent} one infers that the relevant parameter determining the size of the deviations of the truncated Gaussian PDF from 
the standard (parent) Gaussian PDF is given by the positive parameter
\begin{equation}
\label{equ:beta}
{\beta}:=\frac{\mu}{\sqrt{2}\sigma}\;\;.
\end{equation}
The figures \ref{fig3} and \ref{fig4} show that the modes of all histograms are much larger than the associated variances and thus we have $\beta \gg 1$ for the tori considered. 
For $\beta\gg1$, one obtains for $N$ and $\lambda$ from \eref{equ:N} and \eref{equ:lambda}:
\begin{equation}
\label{equ:N2}
{N}=\frac{1}{1-\epsilon}=1+\epsilon+{\mathcal O}(\epsilon^{2}) \ ,
\end{equation}
with
\begin{eqnarray}
\label{equ:epsilon}
{\epsilon}:=\frac{1}{2}\textstyle{\rm erfc}(\beta)=\frac{1}{2\sqrt{\pi}}\frac{e^{-\beta^{2}}}{\beta}\left(1+{\mathcal O}\left(\frac{1}{\beta^{2}}\right) \right) \ ,\quad \textrm{and}\quad
{\lambda}=\frac{1}{\sqrt{2\pi}}e^{-\beta^{2}}+{\mathcal O}\left(\frac{e^{-2\beta^{2}}}{\beta}\right)\; ,\nonumber\\
\end{eqnarray}
which gives e.g. for $\beta=10$ the extremely small values $\epsilon={\mathcal O}(10^{-45})$ and $\lambda={\mathcal O}(\beta\epsilon)$=${\mathcal O}(10^{-44})$.
Thus, it is justified to use for a comparison of the histograms with a Gaussian PDF the Gaussian PDF \eref{equ:pplus} with $N=1$. 
Since a precise determination of the parent parameters $\mu$ and $\sigma$ from the histograms is rendered with difficulties, as discussed before, we show in figures \ref{fig3} and \ref{fig4} the standard Gaussian PDF
determined by the mean values $\langle\rho\rangle$ and the variances $\Sigma^{2}$ of the histograms given in tables \ref{table1} and \ref{table2} for the cases without mask and with U73 mask, respectively.
The ratio $\beta$ is then given by $\langle\rho\rangle / 2\Sigma$, which varies in the unmasked case between $12.340$ and $15.494$, and in the U73 mask case between $12.938$ and $14.532$.
\newpage
\section*{References}
\bibliographystyle{iopart-num}

\end{document}